\documentclass[aps,superscriptaddress,twocolumn,twoside,floatfix,pra,nofootinbib,longbibliography]{revtex4-2}
\usepackage{graphicx,amsmath,amssymb,amsfonts,color}
\usepackage{times,comment,enumerate}
\usepackage[ocgcolorlinks,colorlinks=true,linkcolor=blue,citecolor=red,linktocpage=true,breaklinks=true]{hyperref}
\newcommand{\ket}[1]{|#1\rangle}
\newcommand{\bra}[1]{\langle#1|}
\DeclareMathOperator{\tr}{tr}
\usepackage[framemethod=tikz,linewidth=0.5mm,linecolor=black,backgroundcolor=gray!5,innertopmargin=2mm,innerbottommargin=2mm]{mdframed}

\newcommand{\pid}{\wp}
\newcommand{\epstvd}[1]{\langle\epsilon_{(#1)}\rangle}

\begin{document}

\title{
How to use arbitrary measuring devices to perform almost perfect measurements}

\author{Noah Linden}
\affiliation{School of Mathematics,
University of Bristol,
Fry Building,
Woodland Road, 
Bristol,
BS8 1UG, U.K. }

\author{Paul Skrzypczyk}
\affiliation{H. H. Wills Physics Laboratory, University of Bristol, Tyndall Avenue, Bristol, BS8 1TL, UK.}
\affiliation{CIFAR Azrieli Global Scholars program, CIFAR, Toronto, Canada}

\begin{abstract}
We consider the problem of reproducing one quantum measurement given the ability to perform another. We give a general framework and specific protocols for this problem.  For example, we show how to use available ``imperfect'' devices a small number of times to implement a target measurement with average error that drops off exponentially with the number of imperfect measurements used.   We hope that could be useful in near-term applications as a type of lightweight error mitigation of the measuring devices. As well as the view to practical applications, we consider the question from a general theoretical perspective in the most general setting where both the available and target measurements are arbitrary generalised quantum measurements.  We show that this general problem in fact reduces to the ability to reproduce the statistics of (complete) von Neumann measurements, and that in the asymptotic limit of infinitely many uses of the available measurement, a simple protocol based upon `classical cloning' can perfectly achieve this task. We show that asymptotically all (non-trivial) quantum measurements are equivalent. We also study optimal protocols for a fixed number of uses of the available measurement. This includes, but is not limited to, improving both noisy and lossy quantum measurements. Furthermore, we show that, in a setting where we perform multiple measurements in parallel, we can achieve finite-rate measurement reproduction, by using block-coding techniques from classical information theory. Finally, we show that advantages can also be gained by making use of probabilistic protocols. 

\end{abstract}

\maketitle

\section{Introduction} 
%\textit{Introduction.---}
The process of measurement plays a fundamental role in quantum mechanics, as it is only through measurement that classical information is obtained about the quantum state of a system. The most general measurement that can be performed in principle in quantum mechanics is a so-called generalised measurement (or instrument) -- specified by a set of sub-channels (or usually Kraus operators). %A particularly important class of measurements are so-called ideal von Neumann measurement, whereby to each possible outcome we associate a unit-rank projector, which specifies both the probability of that outcome, via Born's rule, and the post-measurement state. More generally, by bringing in ancillary systems and performing joint measurements on the whole, one can perform more general measurements, known as Positive-Operator-Valued-Measures (POVMs), specified by a set of Kraus operators. 
In practice, in any real experiment it is generally not possible to perform exactly the measurement one wants. For example, when measuring photons, detectors do not always click, leading to losses, or there will be coupling to additional degrees of freedom within the setup. Moreover, in different systems and architectures there may be fundamental or engineering constraints which make certain target measurements impossible or highly demanding to perform. Building high-quality measuring devices is thus a central and demanding task, which has seen significant progress in recent decades as quantum technologies have begun to emerge and mature.  

Here we are interested in approaching the problem of performing a target measurement from the opposite direction: rather than directly engineering it, we are interested in how best to make use of available measuring devices, which may either not be the ones sought after, or which may be imperfect or noisy. In particular, given an available measuring device or devices, we are interested in using it or them in an optimal way, such that at the end of the process, the effective measurement performed is as close as possible to the target measurement. In particular, we are interested in the most general situation, where the available and target measurements are both generalised quantum measurements. The approach we will consider taking is to use the available measuring device \textit{multiple times} -- or equivalent to use multiple identical measuring devices -- in order to realise a \textit{single} use (or even potentially multiple uses) of the target measurement. 

We have both practical and theoretical aims in this paper.

In the context of the former we show how to use available ``imperfect'' devices a few times to implement a target measurement with average error that drops off exponentially with the number of imperfect measurements used.   We hope that this could be useful in near-term applications as a type of lightweight error mitigation of the measuring devices.

At the more abstract theoretical level, we show that all measurements are asymptotically equivalent and that we can achieve finite-rate measurement interconversion, using block-coding techniques from classical information theory.

As an example of the potential types of processes we have in mind, consider that instead of using a measuring device to measure a given system, we first apply a `classical cloning' procedure on the incoming system,
\begin{equation}
	\ket{i} \mapsto \ket{i}\cdots\ket{i},
\end{equation}
e.g. we perform CNOTs with ancillary systems, `cloning' in the standard basis, and spreading information about the incoming state amongst the clones -- and then measure each one of the clones using the available measuring device, before processing the string of outcomes into a single final outcome. What does this final measurement look like, and how does it compare to the available one? As we will show here, this cloning procedure is useful, and can provide lots of flexibility in changing the behaviour of the measuring device, such that it can become close to a target measurement. Crucially however, this procedure is provably not always optimal, and even better protocols can be found, as we will show. 

Although this seems to be a rather natural question, it is one which does not appear to have been addressed in the most general sense considered here in the literature before. The two key papers that have appeared were by Yuen \cite{yuenDesignTransparentOptical1987} who proposed ``purifying'' photodectors using pre-amplification, and by Dall'Arno D'Ariano and Sacchi \cite{dallarnoPurificationNoisyQuantum2010}, who considered the more specific task of purifying noisy measurements. In a completely different direction, QISKIT describe a method of measurement error mitigation based upon linear algebra, post-processing a set of observed counts \cite{Qiskit-Textbook}. A key distinction here is that we are not interested specifically in purification of noisy measurements, but the general question of reproducing a target measurement with an available one, irrespective of how they relate to each other, e.g. even when the measurements have different numbers of outcomes, or are both extremal measurements.

The questions we study here are also interesting from a resource-theoretic perspective \cite{chitambarQuantumResourceTheories2019}. There has been interest lately in applying the resource-theory approach of quantum information to quantum measurements \cite{skrzypczykRobustnessMeasurementDiscrimination2019,oszmaniecOperationalRelevanceResource2019a,takagiGeneralResourceTheories2019a,guffResourceTheoryQuantum2019a}. In this approach, subsets of measurements are considered as the 'free objects', while any other measurement is considered a resource. In this approach, it has however been tacitly assumed that a measuring device is only used once. One main insight of the present paper is that while it is natural in the context of resource theories of quantum states that each state can only be used once for some purpose (e.g. a single ebit which is consumed to perform quantum teleportation \cite{bennettTeleportingUnknownQuantum1993}), given a single measuring device it can normally be used  many times, without being consumed or degraded\footnote{There might be certain situations where this isn't necessarily the case, e.g. a bipartite (LOCC) measurement that consumes entanglement. Here we will focus on situations where it is a reasonable assumption that measuring devices are degraded only very slowly.}. This shows that questions of measurement simulation \cite{gueriniOperationalFrameworkQuantum2017,oszmaniecSimulatingPositiveOperatorValuedMeasures2017} -- which is another way of phrasing the question considered here -- can also naturally consider multiple uses of a measurement. 

In what follows, we will show that in fact, with the ability to perform any given measurement a sufficient number of times, it can approximate any other measurement as closely as desired. We will also show that the optimal procedure for this is non-trivial, in the sense that it is not merely the classical cloning example given above. Then, we consider the asymptotic scenario where we use the measurement many times, and want to approximate numerous uses of a different measurement. We show that we can achieve \emph{finite rate} transformations, using ideas from classical coding theory. Finally, we consider allowing ourselves to make use or probabilistic protocols, and demonstrate that these indeed provide further advantages. We will begin by considering a specific motivating example, which highlights many of the main ideas, before presenting the general results thereafter. 

The paper is organised as follows:
\begin{itemize}
	\item In Sec.~\ref{s:trine} we consider a motivating example, using the `trine' measurement to implement a complete\footnote{Here, by complete, we specifically refer to a von Neumann measurement where each projector is rank-1, and so the measurement has the same number of outcomes as the dimension of the Hilbert space. In contrast, a (non-complete) von Neumann measurement may contain a smaller number of outcomes, with some higher-rank projectors.} von Neumann measurement. These two measurements are very different from each other and thus exhibit many key features of our results.
	In Sec.~\ref{ss:simple protocol} we first consider the two-copy case, and a simple but non-optimal protocol based upon `classical cloning'. In Sec.~\ref{ss:trine general} we then present the general structure of deterministic protocols, before applying this insight back to find optimal (and non-trivial, i.e. not classical cloning) protocols in the case of two-copies in Sec.~\ref{ss:trine 2copy 2},  $N$ copies in Sec.~\ref{ss:trine N copy} and for single copies in Sec.~\ref{ss:trine single copy}. 
	
	\item In section~\ref{ss:noisy Z} we consider the practically important case of imperfect-$Z$ measurements.  An interesting feature is that even here optimal protocols are non-classical cloning. 
	\item In Sec.~\ref{s:gen meas} we move onto the analysis of general measurements. In Sec.~\ref{ss:gen} we outline the general set-up of quantum instruments (generalised measurements). In Sec.~\ref{ss:gen protocols} we then give general deterministic protocols in this setting. In Sec.~\ref{ss:post-meas} we present our `post-measurement' sub-routine, a key building block that implements a generalised measurement given access to a complete\footnote{Note that by `complete' we refer to a von Neumann measurement of a basis, i.e. with rank-1 projectors. This is in contrast to a `non-complete' von Neumann measurement, which would contain higher-rank projectors.} von Neumann measurement. In \ref{ss:asympt} we present our `generalised classical cloning' sub-routine, and present our main result, which combines the two sub-routines to show that any (non-trivial) measurement is able to implement any other measurement in the asymptotic limit of many uses. 
	\item In Sec.~\ref{s:block} we consider the problem of reproducing multiple uses of a target measurement in parallel, and show that in this setting finite rate reproduction is possible, using block-coding protocols from classical information theory. 
	\item In Sec.~\ref{s:prob protocols} we consider probabilistic protocols, and prove that they are able to outperform deterministic protocols. 
	\item Finally, in Sec.~\ref{s:conclusions} we conclude, and discuss future directions of research. 
	
\end{itemize}

%\section{Summary of the main results}

\section{Motivating example: Reproducing a complete von Neumann measurement using the trine}\label{s:trine}
%\textit{Example: Reproducing ideal measurements using the trine.---}
Let us assume that the only available measurement we can perform is the `trine' measurement $\mathbb{M} = \{M_a\}_a$, which has POVM elements 
\begin{equation}
M_a=\frac 23 \ket {\phi_a}\bra{\phi_a}, 
\end{equation}
where
\begin{equation}
\ket {\phi_0} =\ket 0,\quad  \ket {\phi_1}=\frac{\ket 0 +\sqrt 3 \ket 1}{2},\quad\ket {\phi_2}=\frac{\ket 0 -\sqrt 3 \ket 1}{2},
\end{equation}
which are separated by $120^\circ$ in the equatorial plane of the Bloch sphere\footnote{One reason for treating this example in some detail is that, while it may not be particularly relevant in practical applications, many of the key features of the general protocols can be seen in this example.}. Note that we only specify the POVM elements here, as we will assume the measurement is destructive (or alternatively that we are not interested in the post measurement state). In such cases, it is only the POVM elements that are relevant. We will come to the more general case in Sec.~\ref{s:gen meas} below. 

Our goal in this example is to use this measurement $N$ times to reproduce a single (complete) von Neumann measurement $\mathbb{V}~=~\{\Pi_0, \Pi_1\}$ on a qubit with elements $\Pi_0 = \ket{0}\bra{0}$ and  $\Pi_1 = \ket{1}\bra{1}$. Since these two measurements are very different, at first sight this appears to be a challenging task.

In order to formalise our goal more precisely , let us imagine that we are  given a single copy of an unknown state, 
%\begin{equation}
$\ket\psi=\alpha \ket 0 + \beta \ket 1,$
%\end{equation}  
which we are to measure. Our goal then is to reproduce, as closely as possible the complete von Neumann measurement, that is to return outcome $0$ with probability
\begin{equation} \label{e:pid 0}
	\pid(0) = \bra{\psi}\Pi_0\ket{\psi} = |\alpha|^2,
\end{equation} 
and leave the system in the state $\ket{0}$, and to return outcome $1$ with probability 
\begin{equation}\label{e:pid 1}
	\pid(1) = \bra{\psi}\Pi_1\ket{\psi} = |\beta|^2,
\end{equation}
and leave the system in the state $\ket{1}$. We note that if we are able to achieve this goal perfectly, then due to the linearity of quantum mechanics it follows immediately that the correct behaviour will also be reproduced when measuring either a mixed state, or part of a system. That is, if we measured the first particle of an entangled state $\ket{\Phi} = \gamma\ket{0}\ket{\chi_0} + \delta\ket{1}\ket{\chi_1}$, where $\ket{\chi_0}$ and $\ket{\chi_1}$ are normalised, but not necessarily orthogonal states, then we will find $\pid(0) = |\gamma|^2$, $\pid(1) = |\delta|^2$ and the states after measurement $\ket{0}\ket{\chi_0}$ and $\ket{1}\ket{\chi_1}$ respectively, as required. 

In what follows, we will consider protocols which make use of trine measurements to approximate a complete von Neumann measurement. In a protocol that makes use of $N$ trine measurements, we will denote by $P^{(N)}(a)$ the probability that the result $a$ is obtained. $P^{(N)}(a)$ depends upon $\ket{\psi}$, but for notational simplicity, we suppress this dependence. Our goal is that this should be similar to $\pid(a)$, but as we will see, there will be a small state-dependent error in announcing the wrong outcome. 

In order to quantify how `close' our measurement is to the ideal one, we will adopt here the total variation distance (TVD), averaged over all states, 
\begin{multline}
	\epstvd{N} =\int d\psi \Big(\big|P^{(N)}(0)-\pid(0)\big| + \big|P^{(N)}(1)-\pid(1)\big|\Big) .
\end{multline}

We note that in the case here of a dichotomic measurement, we will always have $\pid(1) = 1-\pid(0)$ and $P^{(N)}(1)~=~1-~P^{(N)}(0)$, and so $\epsilon_{(N)}$ simplifies to 
\begin{equation}\label{e:eps N simple}
	\epstvd{N} = 2\int d\psi \big|P^{(N)}(0)-\pid(0)\big| .
\end{equation}

Before presenting our simplest protocol, it will be useful to first consider what can be achieved by just performing the trine measurement, and guessing an outcome for a complete von Neumann measurement based upon the result. The probability distribution when measuring $\ket{\psi} = \alpha \ket{0} + \beta\ket{1}$ is
\begin{align}
P(0) &= \frac{2}{3}|\alpha|^2, \nonumber \\
P(1) &= \frac{1}{6} + \frac{1}{3}|\beta|^2 + \frac{\sqrt 3}{6}(\alpha\beta^* + \alpha^*\beta),\\
P(2) &= \frac{1}{6} + \frac{1}{3}|\beta|^2 - \frac{\sqrt 3}{6}(\alpha\beta^* + \alpha^*\beta) \nonumber
\end{align}
A simple way to approximate the von Neumann measurement is to associate the outcome 0 of the trine measurement with the outcome 0 of the von Neumann measurement, and to associate the other two outcomes of the trine measurement with the outcome 1. In this way, the implemented probabilities are
\begin{align}
    P^{(1)}(0) &= \frac{2}{3}|\alpha|^2, & P^{(1)}(1) &= \frac{1}{3} + \frac{2}{3}|\beta|^2.
\end{align}
Substituting these into \eqref{e:eps N simple}, and evaluating the integral, we can calculate that this simple procedure has average TVD error of
\begin{equation}
    \epstvd{1} = \frac{1}{3}.
\end{equation}

    Maybe unsurprisingly, this error is rather large, and demonstrates that on average, the trine measurement is not able to approximately reproduce the von Neumann measurement under this simple procedure. In the following sections we will see that if we consider more interesting protocols, utilising ancillary particles and making multiple measurements, the situation changes (and indeed these considerations lead us to the realisation that even with one copy of the trine measurement, there are more subtle and better protocols - see section \ref{ss:trine single copy}). 
\subsection{A simple protocol}\label{ss:simple protocol}

\begin{figure}[t!]
	\includegraphics{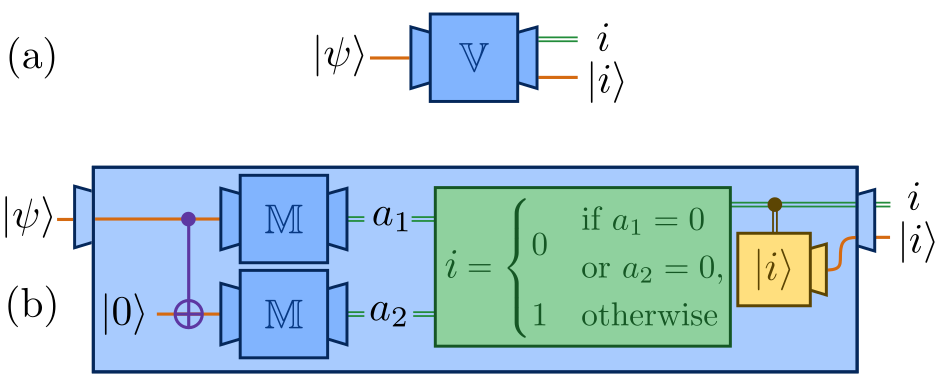}
	\caption{\label{f:two-trine protocol}(a) A complete von Neumann measurement $\mathbb{V} = \{\ket{i}\bra{i}\}$. The probability of the outcome $i$ when measuring the state $\ket{\psi}$ is $\pid(i)~=~|\langle i | \psi\rangle|^2$. The state after measurement is $\ket{i}$. (b) A simple protocol using two copies of a trine measurement to approximately reproduce a complete von Neumann measurement. An ancillary qubit is prepared in the state $\ket{0}$, and the incoming state $\ket{\psi}$ is `classically cloned' by applying a CNOT. Both particles are then measured using the trine measurement, and the outcomes are mapped into the outcome $i$ of the measurement. Conditional on the outcome $i$, the corresponding state $\ket{i}$ is prepared. }
	
\end{figure}

We will exhibit first a simple protocol, which makes use of the trine measurement twice in order to implement a measurement which is not too dissimilar from the complete von Neumann measurement, that demonstrates many of the key features we are interested in. The basic intuition is to spread out the information contained in the state (the amplitudes $\alpha$ and $\beta$), into a larger Hilbert space, before attempting to retrieve the information using multiple measurements. This protocol is summarised pictorially in Fig.~\ref{f:two-trine protocol} (b). We will see shortly however that this simple protocol is not optimal. 

Let us consider that we prepare an ancillary system in the state $\ket{0}$, and `classically clone' -- i.e.~apply a controlled-not unitary between this system and $\ket{\psi}$, with $\ket{0}$ as the target, leading to the (in general entangled) state
\begin{equation}
	\ket{\Psi} = \alpha \ket{0}\ket{0} + \beta \ket{1}\ket{1}.
\end{equation}
We now measure both of the qubits, each one with the trine measurement. The probability distribution for the pairs of outcomes is found to be
\begin{align}\label{e:probs}
	P(0,0) &= \frac{4}{9}|\alpha|^2,\nonumber \\
	P(0,1) &= P(0,2)
	= P(1,0) = P(2,0) = \frac{1}{9}|\alpha|^2, \\
	P(1,1)  &= P(2,2) = \frac{1}{36} + \frac{2}{9}|\beta|^2 + \frac{1}{12}(\alpha\beta^* + \alpha^*\beta), \nonumber \\
	P(1,2)  &= P(2,1) = \frac{1}{36} + \frac{2}{9}|\beta|^2 - \frac{1}{12}(\alpha\beta^* + \alpha^*\beta), \nonumber
\end{align}
where $P(a_1, a_2) = \bra{\Psi}M_{a_1}\otimes M_{a_2}\ket{\Psi}$ is the joint probability for the pair of outcomes $(a_1,a_2)$. In order to try and implement the complete von Neumann measurement, we need to map these 9 outcomes-pairs onto the two outcomes $0$ and $1$ of the target measurement, which is what will be declared as the result of the measurement. A choice which works well is 
\begin{align}\label{e:optimal deterministic trine}
	\{(0,0), (0,1), (0,2), (1,0), (2,0)\} &\mapsto 0,\nonumber \\
	\{(1,1), (1,2), (2,1), (2,2)\} &\mapsto 1, 
\end{align}
that is, in the first five cases from \eqref{e:probs}, where at least one measurement returns a zero outcome are mapped to $0$, while the final four cases from \eqref{e:probs}, where neither measurement returns a zero outcome are mapped to 1. Under this mapping, the implemented probabilities $P^{(2)}(i)$ (where the superscript $(2)$ reminds us that we are considering two uses of the trine) are seen to be
\begin{align}
	P^{(2)}(0) &= \frac{8}{9}|\alpha|^2,\nonumber \\
	P^{(2)}(1) &= \frac{1}{9} + \frac{8}{9}|\beta|^2.
\end{align}

Given these probabilities, it is straightforward to calculate the average TVD error $\epsilon_{(2)}$, which is found to be
\begin{equation}
	\epstvd{2} = \frac{1}{9}. 
\end{equation}
Only in the case that the probabilities $P^{(2)}(i)$ were identical to the target probabilities $\pid(i)$ for all states $\ket{\psi}$ would $\epstvd{2}$ equal zero. The small value found here shows that, on average, the error in the probabilities is indeed small, as expected.

As we explain in more detail below, we can also calculate the POVM $\mathbb{M}' = \{M'_i\}_i$ that the above procedure implements. We find 
\begin{align}
	M'_0 &= \frac{8}{9}\ket{0}\bra{0},& M'_1 &= \frac{1}{9}\ket{0}\bra{0} + \ket{1}\bra{1}.
\end{align}

We also want to leave the system in the correct state after the measurement. This step is however straightforward to achieve. Whenever we announce the result $i$, we simply prepare the system in the state $\ket{i}$. 

Altogether, this simple example, making use of the trine measurement only twice, demonstrates that we can in fact implement a measurement that is not far from the complete von Neumann measurement, even though on the face of it, they are decidedly different measurements. 

\subsection{More general protocols}\label{ss:trine general}

\begin{figure}[t!]
	\includegraphics{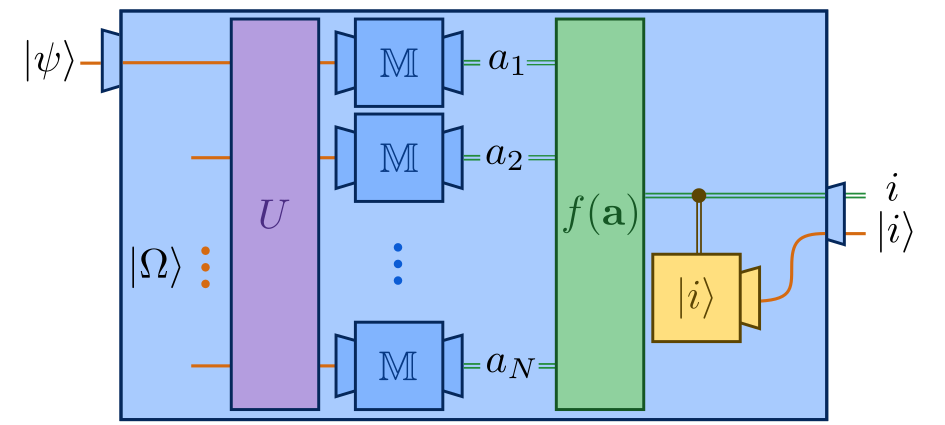}
	\caption{\label{f:trine protocol} Protocols using $N$ copies of the trine measurement in order to approximately reproduce a complete von Neumann measurement. $N-1$ ancillary qubits are prepared in the state $\ket{\Omega}$ and a global unitary $U$ is applied to these as well as the state $\ket{\psi}$. All $N$ qubits are then measured using the trine, and the string of outcomes $\mathbf{a} = (a_1,\ldots,a_N)$ is mapped into an outcome $i = f(\mathbf{a})$. Conditional on $i$, the state $\ket{i}$ is prepared. This implements a measurement $\mathbb{M}^{(N)}$.}

\end{figure}
We will now show that using the available measurement multiple times allows an exponential decrease of the average error in the number of uses.

We consider a general class of protocols of the above type, where we are now able to perform trine measurements $N$ times, and want to reproduce as closely as possible a single von Neumann measurement. The class of protocols we will consider are summarised pictorially in Fig.~\ref{f:trine protocol} and will comprise the following elements: 
\begin{enumerate}[(i)]
	\item We prepare $N-1$  ancillary qubits in a state $\ket\Omega$, independent of $\ket\psi$. Without loss of generality (because of the next step), this can be taken to be the tensor product of $N-1$ copies of $\ket 0$. As such, we are considering a collection of $N$ qubits. 
	\item We perform a global unitary transformation $U$ on these $N$ qubits, such that the state becomes 
\begin{equation}
\ket \Psi = U\ket\psi\ket\Omega = \alpha \ket{\Psi_0} + \beta \ket{\Psi_1},
\end{equation}
where  
\begin{align}
\ket{\Psi_0}&=U\ket 0\ket\Omega,& \ket{\Psi_1}&=U\ket 1\ket\Omega, 
\end{align}
are two orthogonal  and normalised, but otherwise arbitrary states of the $N$ qubits.
\item We measure all $N$ systems, using the trine measurement.\footnote{We will see below, that in fact this is not quite as general as we need; we will also need in special cases to only measure a \emph{subset} of the systems, and trace out the remaining, unmeasured systems.}
\item Dependent on the string $\mathbf{a} = (a_1,a_2,\cdots,a_N)$ of outcomes of the $N$ measurements, we return as outcome of the target measurement either $0$ or $1$, and prepare the corresponding state $\ket 0$ or $\ket 1$.
\end{enumerate}

There are two approaches that can be taken in the final step, either a deterministic or a probabilistic assignment. In the former case, to each string of outcomes we associate a definite outcome, either $0$ or $1$. We can do this, for example, by specifying a function $f(\mathbf{a})$, such that we assign the value $i = f(\mathbf{a})$ to the string of outcomes $\mathbf{a}$, as we did earlier in the two-copy case. In the latter case, on the other hand, this assignment is allowed to be probabilistic. This is then specified instead by a conditional probability distribution $q(i|\mathbf{a})$, which gives the probability of assigning the outcome $i$ to the string $\mathbf{a}$.

In what follows we will restrict our attention to deterministic protocols.  We will return to probabilistic protocols later on in Sec.~\ref{s:prob protocols}, where we will show that, surprisingly, they can offer an advantage over deterministic protocols, at the expense of being much harder to analyse in general. 

In either of these two cases, the above procedure defines a POVM $\mathbb{Q}^{(N)}$ with elements $Q^{(N)}_0$ and $Q^{(N)}_1 = (\mathbb{I}_{N}-Q^{(N)}_0)$ on the $N$ systems, associated to the outcomes 0 and 1 respectively, where $\mathbb{I}_n$ denotes the identity on $n$ qubits. In particular, in the deterministic case, these are given by
\begin{align}
	Q^{(N)}_i = \sum_{a_1,\cdots,a_N} \delta_{i,f(\mathbf{a})} M_{a_1} \otimes \cdots \otimes M_{a_N},
\end{align}
That is, $Q^{(N)}_0$ and $Q^{(N)}_1$ are sums of tensor products of trine POVM elements. The implemented probabilities are then given by
\begin{align}\label{e:PN}
	P^{(N)}(i) &= \bra{\Psi}Q^{(N)}_i\ket{\Psi},\nonumber \\
	&= \bra{\psi}\bra{\Omega}U^\dagger Q^{(N)}_i U \ket{\psi}\ket{\Omega},
\end{align}
from which it follows that the associated implemented POVM elements are
\begin{equation}\label{e:sim POVM}
	M^{(N)}_i = \tr_{2\cdots N}[(\mathbb{I}\otimes \ket{\Omega}\bra{\Omega})U^\dagger Q_i^{(N)} U],
\end{equation}
where $\tr_{2\cdots N}$ denotes the partial trace over all but the first system. In what follows, we will find it useful to think about properties of both the POVM $\mathbb{Q}^{(N)}$ acting on all $N$ systems and in the effective POVM $\mathbb{M}^{(N)}$ acting on the system alone.

The figure of merit we consider will remain the same, as given in \eqref{e:eps N simple}. In the Appendix we show that the average (integral) over all states can be performed -- under the natural assumption that the implemented POVM elements are diagonal in the basis of the target von Neumann measurement\footnote{This can in fact be achieved without any loss of generality -- if we have a protocol which doesn't have this property, we can always `dephase' it (making it diagonal in the target basis) by interacting with one further ancillary system. We have not found any advantage in considering protocols which produce coherence in the target basis}. Under this assumption, the implemented POVM element $M_0^{(N)}$ has the form 
\begin{equation}
    M_0^{(N)} = (1-\epsilon_0)\ket{0}\bra{0} + \epsilon_1 \ket{1}\bra{1},
\end{equation}
where 
\begin{align}\label{e:x and y}
	\epsilon_0 &= \bra{\Psi_0}Q^{(N)}_1\ket{\Psi_0} = 1-\bra{\Psi_0}Q^{(N)}_0\ket{\Psi_0},\\
	\epsilon_1 &= \bra{\Psi_1}Q^{(N)}_0 \ket{\Psi_1},
\end{align}
are the errors, respectively, that the outcome 1 is returned when the state is $\ket{0}$ and that 0 is returned when the state is $\ket{1}$, and we used the fact that $Q_1^{(N)} = \mathbb{I} - Q_0^{(N)}$. Thus, using the standard parametrisation for the state $\ket\psi~=~\alpha\ket 0 ~+~\beta\ket 1~=~\cos \frac{\theta}{2} \ket 0 + e^{i\phi}\sin\frac{\theta}{2}\ket 1$, it follows that $P^{(N)}(0) = (1-\epsilon_0)\cos^2 \frac{\theta}{2} + \epsilon_1 \sin^2 \frac{\theta}{2}$. Substituting this into \eqref{e:eps N simple}, we are then able to explicitly integrate, after identifying the two regions where the $P^{(N)}(0)~-~\pid(0)$ has differing sign. A straightforward but lengthy calculation then leads to the following simplified form 
\begin{equation}
	\epstvd{N} = \frac{\epsilon_0^2 +\epsilon_1^2}{ \epsilon_0 + \epsilon_1} \label{quadratic1} 
\end{equation}
Given \eqref{e:x and y}, $1-\epsilon_0$ and $\epsilon_1$ must lie between the minimal and maximal eigenvalues of $Q_0^{(N)}$, but are otherwise unconstrained. This now provides the possibility of finding the \emph{best} possible average TVD error, by performing the simple optimisation
\begin{align}
	\epstvd{N}^* = \min_{\epsilon_0,\epsilon_1}& \quad \frac{\epsilon_0^2 +\epsilon_1^2}{\epsilon_0 + \epsilon_1} \label{quadratic} \\
	\text{s.t. }& \quad 1-\lambda_{\rm max}\leq \epsilon_0 \leq1-\lambda_{\rm min}, \nonumber\\
	& \quad \lambda_{\rm min}\leq \epsilon_1 \leq\lambda_{\rm max}, \nonumber
\end{align}
where $\lambda_{\rm min}$ and $\lambda_{\rm max}$ are the minimum and maximum eigenvalues of $Q_0^{(N)}$, which must each lie in the interval $[0,1]$.

\begin{table*}[t]
	\centering
	{\renewcommand{\arraystretch}{2}\setlength{\tabcolsep}{1em}
		\begin{tabular}{c|c|c}
			Region & $(\epsilon_0^*, \epsilon_1^*)$ & $\epstvd{N}^*$ \\ \hline
			$\lambda_{\rm min} < (\sqrt{2}-1)(1-\lambda_{\rm max})$ & $\left(1-\lambda_{\rm max}, (\sqrt{2}-1)(1-\lambda_{\rm max})\right)$ & $2(\sqrt{2}-1)(1-\lambda_{\rm max})$ \\
			$(\sqrt{2}-1)(1-\lambda_{\rm max}) \leq \lambda_{\rm min} \leq (\sqrt{2}+1)(1-\lambda_{\rm max})$ & $(1-\lambda_{\rm max},\lambda_{\rm min})$ & $\displaystyle{\frac{(1-\lambda_{\rm max})^2 + \lambda_{\rm min}^2}{1-\lambda_{\rm max} + \lambda_{\rm min}}}$ \\ 
			$(\sqrt{2}+1)(1-\lambda_{\rm max}) < \lambda_{\rm min}$ & $\left((\sqrt{2}-1)\lambda_{\rm min},\lambda_{\rm min}\right)$ & $2(\sqrt{2}-1)\lambda_{\rm min}$
	\end{tabular}}
	\caption{Regions and behaviour of $\epstvd{N}^*$. As shown in the Appendix, there are three regions, depending upon the relation between $\lambda_{\rm min}$ and $\lambda_{\rm max}$. In each of these regions, it is possible to analytically find the minimising pair $(\epsilon_0^*, \epsilon_1^*)$ of \eqref{quadratic}, and the corresponding optimal average TVD error $\epstvd{N}^*$. \label{tab:1}}
\end{table*}

In the Appendix, using the structure of \eqref{quadratic} and identifying the geometry of the problem, we show that it is possible to analytically solve for the optimal $\epsilon_0^*$ and $\epsilon_1^*$ that achieve the minimum in \eqref{quadratic}. The results are summarised in Table.~\ref{tab:1}. 

This shows that it is in fact straightforward to solve for the minimum average TVD error $\epstvd{N}^*$ once a partition of the outcome strings has been made. Indeed, once we have chosen a partition, this defines the POVM $\mathbb{Q}^{(N)}$, which determines $\lambda_{\max}$ and $\lambda_{\min}$. Moreover, once we have solved \eqref{quadratic}, we can finalise the protocol, since from \eqref{e:x and y} we see that we just need to find two orthogonal states $\ket{\Psi_0}$ and $\ket{\Psi_1}$ that have the correct expectation values on $Q_0^{(N)}$, and we can always do this.\footnote{As we will see later, while it is simple to achieve the correct expectation values, it is not so obvious at first sight how to ensure that the states are orthogonal. This can however always be achieved by making use of extra ancillary particles, which are not measured.}  Thus, we see that we have reduced our initial problem to a much simpler one, of solving a simple quadratic problem. The final stage is then to find the optimal partitions, which needs to be done either by appealing to a careful argument, or by simply checking the (large) number of possibilities.

There are furthermore a number of interesting lessons we can draw from this reformulation of $\epstvd{N}$. First, we can notice that it only depends upon two specific states, $\ket{\psi} = \ket{0}$ and $\ket{\psi} = \ket{1}$, through $\ket{\Psi_0}$ and $\ket{\Psi_1}$. These are the two orthogonal states that the von Neumann measurement is able to perfectly distinguish. As such, we see that the question of how well our implemented measurement performs on the entire Bloch sphere is reduced to a modified figure of merit on only two natural states.

Second, we note that from the form \eqref{quadratic1}  it is straightforward to see, since both terms in the numerator are non-negative,  and since the  denominator cannot become negative when $0 \leq \epsilon_0,\epsilon_1 \leq 1$, that the only way that the average TVD error $\epstvd{N}$ can vanish is if $\epsilon_0 = 0$ and $\epsilon_1 = 0$. This is of course as we should have expected -- it corresponds to the case whereupon receiving $\ket{i}$ as the state to be measured, the measurement returns $i$ with certainty. Crucially however, when considering finding the optimal error, due to the constraints in \eqref{quadratic} this in turn means that $\lambda_{\min} = 0$ and $\lambda_{\max} = 1$. This further means that $Q_0^{(N)}$ must have an eigenvalue $0$ and an eigenvalue $1$. The former condition, due to the fact that $Q_1^{(N)} = \mathbb{I} - Q_0^{(N)}$ means that $Q_1^{(N)}$ must have an eigenvalue equal to 1. Altogether therefore this shows that in order to have zero error, the POVM $\mathbb{Q}^{(N)}$ must be such that each of its elements has a unit eigenvalue, and then it is clear that we must arrange things such that $\ket{\Psi_0} = U\ket{0}\ket{\Omega}$ and $\ket{\Psi_1} = U\ket{1}\ket{\Omega}$ are the corresponding eigenvectors. 

Conversely, the above shows that if  $\lambda_{\min} > 0$ or $\lambda_{\max} < 1$, we will for certain have $\epstvd{N}^* > 0$, corresponding to an imperfect reproduction of the target complete von Neumann measurement. Therefore, our problem has reduced to the study of the eigenvalues of the POVM $\mathbb{Q}^{(N)}$, and in particular how close to 1 we can arrange the largest eigenvalue of $Q_0^{(N)}$ to be. 

\subsection{Returning to two-copies}\label{ss:trine 2copy 2}

Let us return to our example from Sec.~\ref{ss:simple protocol}, using two copies of the trine, $N=2$, and rephrase everything in the above terms. The strategy that we outlined previously amounts to choosing
\begin{align}\label{e:Q2}
	Q_0^{(2)} &= M_0 \otimes M_0 + M_0 \otimes M_1 + M_0 \otimes M_2 \nonumber \\
	&\quad\quad\quad + M_1 \otimes M_0 + M_2 \otimes M_0\nonumber \\
	&= \mathbb{I}\otimes\mathbb{I} - (\mathbb{I}-M_0)\otimes(\mathbb{I}-M_0),\\
	Q_1^{(2)} &= M_1 \otimes M_1 + M_1 \otimes M_2 + M_2 \otimes M_1  + M_2 \otimes M_2 \nonumber \\
	&=  (\mathbb{I}-M_0)\otimes(\mathbb{I}-M_0).\nonumber 
\end{align}
for the POVM $\mathbb{Q}^{(2)}$, and $U$ equal to controlled not such that 
\begin{align}
	\ket{\Psi_0} &= \ket{0}\ket{0},& \ket{\Psi_1} &= \ket{1}\ket{1},
\end{align}
and 
\begin{align}\label{e:guess P2}
	P^{(2)}(0) &= \frac{8}{9}\text{ when }\ket{\psi} = \ket{0}\nonumber,\\ P^{(2)}(0) &= 0 \text{ when }\ket{\psi} = \ket{1},
\end{align}
which led, as a reminder, to $\epstvd{2} = \frac{1}{9}$.

Now, $Q_0^{(2)}$ has minimum and maximum eigenvalues equal to $\lambda_{\min} = 0$ and $\lambda_{\max} = \frac{8}{9}$ respectively. From Table~\ref{tab:1} we can verify we are in the first case, and that the average TVD error is thus
\begin{equation}
\epstvd{2}^* = \frac{2(\sqrt{2}-1)}{9} \approx 0.092
\end{equation}
which is achieved at
\begin{equation}
	(\epsilon_0^*,\epsilon_1^*) = (\tfrac{1}{9}, \tfrac{\sqrt{2}-1}{9})
\end{equation}
and is strictly better than what we found before, which we see is $\frac{1}{2(\sqrt{2}-1)} \approx 1.21$ times larger than the optimal average TVD error. We thus see that the classical cloning protocol we gave in Sec.~\ref{ss:simple protocol}, although intuitive, isn't in fact the optimal strategy. It will be very insightful now to understand why there is a better strategy, and how it differs from our initial guess. 

To that end, we can notice, crucially, that while $\epsilon_0^* = 1-\lambda_{\max}$, on the contrary the optimal value $\epsilon_1^* > \lambda_{\min} = 0$. That is, in contrast to the procedure from before, which led to the probabilities given in \eqref{e:guess P2}, the optimal probabilities are in fact
\begin{align}\label{e:best P2}
	P^{(2)*}(0) &= \frac{8}{9} \text{ when } \ket{\psi} = \ket{0}\nonumber \\
	P^{(2)*}(0) &= \frac{\sqrt{2}-1}{9} \text{ when } \ket{\psi} = \ket{1},
\end{align}
and the optimal implemented POVM is
\begin{align}
	M_0^{(2)*} &= \frac{8}{9}\ket{0}\bra{0} + \frac{\sqrt{2}-1}{9}\ket{1}\bra{1},\nonumber \\  M_1^{(2)*} &= \frac{10-\sqrt{2}}{9}\ket{1}\bra{1} + \frac{1}{9}\ket{0}\bra{0}.
\end{align}
This shows that it is not optimal to take  $\bra{\Psi_1}Q_0^{(2)}\ket{\Psi_1}$ to be equal to the smallest eigenvalue of $Q_0^{(2)}$, and furthermore that in general $\ket{\Psi_1}$ will not be an eigenstate of $Q_0^{(2)}$. This can be seen explicitly in this example -- the eigenvalues of $Q_0^{(2)}$ are $\{\frac{8}{9}, \frac{2}{3}, \frac{2}{3}, 0\}$ with corresponding eigenvectors $\{\ket{0}\ket{0}, \ket{0}\ket{1},\ket{1}\ket{0}, \ket{1}\ket{1}\}$, and so we cannot take $\ket{\Psi_1^*}$ to be an eigenstate of $Q_0^{(2)}$ and achieve $\epsilon_1^*$.   

Since $\epsilon_0^* = 1-\lambda_{\max}$, we must take $\ket{\Psi_0^*} = \ket{0}\ket{0}$, the corresponding eigenvector, just as before. For $\ket{\Psi_1^*}$ we see that we have some freedom, showing that there are infinitely many optimal solutions. One particularly simple choice is
\begin{equation}
	\ket{\Psi_1^*} = \ket{1}\left(\sqrt{\frac{\sqrt{2}-1}{6}}\ket{0} + \sqrt{\frac{7-\sqrt{2}}{6}}\ket{1}\right),
\end{equation}
which is still achieved by $U$ being a controlled unitary transformation, except now when the control is $\ket{1}$ the unitary applied is no longer a full $X$ rotation, but a rotation by a slightly smaller angle. 

At first sight, it might appear that this protocol should be worse than the previous protocol, since before if we were given the state $\ket{\psi} = \ket{1}$, then there was zero probability of erroneously returning the result 0, whereas with this new strategy the probability is $\frac{\sqrt{2}-1}{9}$. However, if we consider a state on the equator of the Bloch sphere, $\ket{\psi} = (\ket{0} + e^{i\phi}\ket{1})/\sqrt{2}$, then the previous protocol gave $P^{(2)}(0) = \frac{4}{9} \approx 0.44$, while the optimal protocol gives $P^{(2)*}(0) = \frac{7+\sqrt{2}}{18} \approx 0.47$, independent of $\phi$, which is closer to the target statistics of $\pid(0) = \frac{1}{2}$. The lesson here is that looking only at the statistics of the basis states can be misleading compared to the statistics on average over the entire Bloch sphere, and this averaging is precisely what is taken into account in order to arrive at the form \eqref{quadratic}.

Finally, we should note that in the above we have found the optimal unitary $U$, and hence optimal states $\ket{\Psi_0^*}$ and $\ket{\Psi_1^*}$ after having fixed a choice of $\mathbb{Q}^{(2)}$. This raises the possibility that there is in fact a better choice of $\mathbb{Q}^{(2)}$, i.e.~a better partition of the strings $\mathbf{a}$ into two subsets. An exhaustive search finds this not to be the case; the protocol presented is in fact an optimal choice of $\mathbb{Q}^{(2)}$ (other optimal choices also exist, which relate to this choice by suitable symmetries).

\subsection{$N$-copy protocols}\label{ss:trine N copy}

We now briefly outline how the above generalises to the case of $N$ uses of the trine. The natural generalisation is  
\begin{align}
	Q_1^{(N)} &=  (M_1 + M_2)^{\otimes N} \equiv (\mathbb{I}-M_0)^{\otimes N},\nonumber \\
	Q_0^{(N)} &= \mathbb{I}_N - (M_1 + M_2)^{\otimes N} \equiv \mathbb{I}_N - (\mathbb{I}-M_0)^{\otimes N}.
\end{align}
Exhaustive search confirms this to be an optimal choice amongst all deterministic protocols for two- and three-copy protocols.\footnote{We note that this is not the unique optimal protocol; other optimal protocols are related to this one by symmetry.} The generalisation to $N$ copies corresponds to outputting $1$ if none of the trine measurements have as outcome $0$, and returning $0$ in the other case, when one or more trine measurement returns $0$.   

$Q_0^{(N)}$ has minimum and maximum eigenvalues equal to 
\begin{align}
\lambda_{\min} &= 0,&  \lambda_{\max} &= 1 - \frac{1}{3^N}. 
\end{align}
Making use of Table~\ref{tab:1}, we remain in the 1st case, and have 
\begin{equation}
	\epstvd{N}^* = \frac{2(\sqrt{2}-1)}{3^N},
\end{equation}
which is achieved at
\begin{equation}
	(\epsilon_0^*,\epsilon_1^*) = \left(\frac{1}{3^N}, \frac{\sqrt{2}-1}{3^N}\right). 
\end{equation}
Crucially, this shows that the average TVD error drops off exponentially fast in the number of uses of the trine, and, surprisingly, tends to a perfect reproduction in the asymptotic limit of infinitely many uses. Thus, even though the trine measurement is fundamentally distinct from the complete von Neumann measurement, if we allow ourselves the possibility of performing the measurement many times, it in fact becomes equivalent. 

An optimal unitary $U$ achieving this protocol\footnote{As in the previous example, there are infinitely many unitaries we can choose between; here we focus on what we believe is the simplest possible protocol to achieve in the lab.} produces
\begin{align}\label{e:U_N}
	\ket{\Psi_0} &= U\ket{0}^{\otimes N} = \ket{0}^{\otimes N},\nonumber \\
	\ket{\Psi_1} &= U\ket{1}\ket{0}^{\otimes (N-1)} = \ket{1}^{\otimes (N-1)}\ket{\phi^{(N)}},\nonumber \\
\end{align}
with
\begin{equation}
\ket{\phi^{(N)}} = 	\sqrt{\frac{\sqrt{2}-1}{2\cdot 3^{N-1}}}\ket{0} + \sqrt{1-\frac{\sqrt{2}-1}{2 \cdot 3^{N-1}}}\ket{1}.
\end{equation}
That is, the optimal protocol is very close to a `classical cloning' protocol, whereby we copy $\ket{0}$ to $N$ copies of $\ket{0}$, and $\ket{1}$ to $N$ copies of $\ket{1}$. The key difference is that the final qubit should not be copied exactly, but must be rotated toward $\ket{1}$, with distance decreasing exponentially with $N$. A simple calculation shows that if we instead did use the classical cloning protocol, this would amount to picking the suboptimal pair $(\epsilon_0,\epsilon_1) = \left(\frac{1}{3^N}, 0\right) = (1-\lambda_{\max},\lambda_{\min})$, which would achieve $\epstvd{N} = \frac{1}{3^N}$. Notably this is always $\frac{1}{2(\sqrt{2}-1)} \approx 1.21$ times worse in average TVD error compared to the optimal choice, independent of $N$, and therefore still has an error which drops off exponentially fast to zero. 

The measurement we implement through this $N$-use protocol, as given by \eqref{e:sim POVM} is found to be
\begin{align}\label{e:M_N}
	M_0^{(N)} &= \left(1-\frac{1}{3^N}\right)\ket{0}\bra{0} + \frac{\sqrt{2}-1}{3^N}\ket{1}\bra{1},\nonumber \\  
	M_1^{(N)} &= \left(1-\frac{\sqrt{2}-1}{3^N}\right)\ket{1}\bra{1} + \frac{1}{3^N}\ket{0}\bra{0}.
\end{align}

Before moving onto the case of more general measurement implementations, there are two final points worth discussing. 
\subsection{Single-use protocol}\label{ss:trine single copy}
The first point worth noting is that our procedure shows that already in the scenario where the trine measurement is performed only once there is a non-trivial protocol to follow. In particular, substituting $N = 1$ into \eqref{e:M_N} indicates we should be able to achieve an implementation of the noisy complete von Neumann measurement
\begin{align}
	M_0^{(1)} &= \frac{2}{3}\ket{0}\bra{0} + \frac{\sqrt{2}-1}{3}\ket{1}\bra{1},\\ M_1^{(1)} &= \frac{4-\sqrt{2}}{3}\ket{1}\bra{1} + \frac{1}{3}\ket{0}\bra{0}, 
\end{align}
which achieves an average TVD error of $\epstvd{1}^*~=~\frac{2}{3}(\sqrt{2}~-~1) \approx 0.276$. Notably, this outperforms the `naive' procedure of just measuring using the trine and outputting $0$ when the outcome is 0 -- the procedure that we started with at the beginning of Sec.~\ref{s:trine}. We saw previously that the naive procedure achieves $\epstvd{1} = \frac{1}{3}$.

\begin{figure}[t!]
	\includegraphics{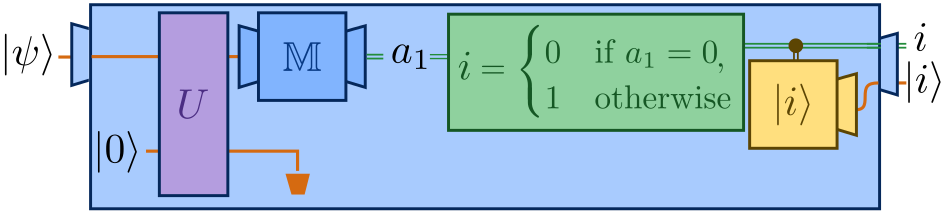}
	\caption{\label{f:single-trine protocol}A modified protocol in order to optimally make use of a single trine measurement. This protocol makes use of an ancillary qubit, so that the unitary $U$ from \eqref{e:U single trine} can be applied, which effectively allows the states $\ket{\Psi_0}$ and $\ket{\Psi_1}$ to be non-orthogonal. Since only a single trine measurement is performed, the ancillary particle is left unmeasured / is thrown away at the end of the protocol.  }
	
\end{figure}

We do however appear to run into a difficulty when we consider the unitary $U$ that we need to perform from \eqref{e:U_N}, since this is in fact no longer unitary, taking orthogonal states to non-orthogonal states. There is however an avenue to overcome this small difficulty, as depicted in Fig.~\ref{f:single-trine protocol} Consider preparing an ancillary qubit in the state $\ket{0}$, which is \emph{not} measured. This ancillary qubit allows us to enact the transformation we need on the first qubit, and flip the second so as to satisfy the orthogonality requirements, i.e.
\begin{align}\label{e:U single trine}
	U\ket{0}\ket{0} &= \ket{0}\ket{0},\\
	U\ket{1}\ket{0} &= \left(\sqrt{\frac{\sqrt{2}-1}{2}}\ket{0} + \sqrt{\frac{3-\sqrt{2}}{2}}\ket{1}\right)\ket{1}.
\end{align}
 It is interesting that it is necessary to go beyond the use of a single qubit -- which effectively allows us to move from unitary transformations to more general CPTP maps at the level of the first qubit -- and that this is useful.  

\section{Imperfect $Z$ measurements}\label{ss:noisy Z}
We now turn to the case that our available physical devices perform an imperfect version of a (complete) von Neumann measurement.  This case is similar to one that occurs in many physically important applications.

Interestingly, this example also sheds light on the question of why is it that protocols become non-trivial, requiring us to move beyond simple classical cloning.  A natural conjecture would be that this derives from the fact that the trine measurement contains three non-orthogonal rank-one elements, which do not commute with each other. Maybe it is this non-commutativity which leads to the interesting structure to the problem we identified above. In what follows, we will show that this is not the case, and that this is not the origin of the non-trivial aspects of our measurement implementation problem. 

Consider a simple class of measurements with POVM elements
\begin{align}
	N_0(p,q) &= p \ket{0}\bra{0} + (1-q)\ket{1}\bra{1},\nonumber \\
	 N_1(p,q) &= (1-p)\ket{0}\bra{0} + q\ket{1}\bra{1}.
\end{align}
w.l.o.g. we take $p>q$. We can view $p,q \in (\frac{1}{2},1]$ and  as parametrising the asymmetric noise\footnote{Note that we restrict the interval for later convenience, as it will be sufficient for our purposes here.} of the measurement. We can think of this class of measurements as being asymmetric noisy $Z$ measurements, which include the standard noisy $Z$ measurement in the special symmetric case $p = q$. 

We will consider the situation of a single-copy ($N = 1$) approximate implementation of the von Neumann measurement using these noisy measurements. In this case, we have $\lambda_{\min} = 1-q < \frac{1}{2}$ and $\lambda_{\max} = p > \frac{1}{2}$.

From Table~\ref{tab:1}, we know that there are three regions to consider. In the middle region, where
\begin{equation}\label{e:p q regions}
(\sqrt{2}+1)q - \sqrt{2} \leq p \leq 2-\sqrt{2} + (\sqrt{2}-1)q,
\end{equation}
 the optimal solution is obtained at $(\epsilon_0^*, \epsilon_1^*) = (1-\lambda_{\max},\lambda_{\min})$. This regime includes the standard noisy $Z$ measurement. In this regime an optimal protocol is just the trivial (classical cloning) one. 

Surprisingly, the remaining two regions, where there is a non-trivial (not classical cloning) protocol to follow, are in fact non-empty. In particular, when either 
\begin{align}
	p &> 2-\sqrt{2} + (\sqrt{2}-1)q& &\text{or}& p< (\sqrt{2}+1)q - \sqrt{2},
\end{align}
i.e. when we consider a sufficiently asymmetric situation, this is the case. Focusing on the former case, Table~\ref{tab:1} shows that the optimal solution is at 
\begin{equation}
(\epsilon_0^*,\epsilon_1^*) = \left((\sqrt{2}-1)(1-q),1-q\right), 
\end{equation}
which achieves $\epstvd{1}^* = 2(\sqrt{2}-1)(1-q)$, and for which an optimal protocol, which requires an ancillary qubit (which will not be measured, as in the case of a single use of the trine), is to perform the transformation
\begin{align}
	U\ket{0}\ket{0} &= (\gamma\ket{0} + \delta \ket{1})\ket{0},& U\ket{1}\ket{1} = \ket{1}\ket{1},
\end{align}
where $\delta = \sqrt{1-\gamma^2}$, and $\gamma$ depends upon $p$ and $q$, and is equal to
\begin{equation}
	\gamma = \sqrt{\frac{1-\sqrt{2} + \sqrt{2}q}{p+q-1}}.
\end{equation}
One can confirm that $0 \leq \gamma \leq 1$ for all $p$ and $q$ in the first regime in \eqref{e:p q regions}. The protocol is to measure the first of the two qubits, and return as the outcome the observed outcome (and prepare the corresponding state). 

What this shows is that even though we have a very simple measurement, with two commuting POVM elements, there is a regime where it is nevertheless necessary to utilise a non-trivial (not classical-cloning) protocol in order to use it to optimally implement a complete von Neumann measurement. Thus, the need for non-trivial protocols does not derive from the non-commutativity of the POVM elements, and is a general feature of optimal protocols.

\section{Reproducing general measurements}\label{s:gen meas}
Up until this point we focused on a particularly insightful problem -- that of reproducing a single use of a qubit von Neumann measurement, given the ability to perform the trine measurement $N$ times. In this section we will now study the more general problem, of reproducing the single use of one measurement given the ability to perform another measurement $N$ times. The main result of this section will be to show that in fact any measurement can reproduce any other measurement -- showing that all measurements are in fact equivalent in the setting we consider. Crucially, this is true even when the measurement we are trying to reproduce has non-trivial post-measurement states (in contrast to the complete von Neumann measurement, as we will explain more carefully below). A key insight is that in fact the ability to reproduce a complete von Neumann measurement can be viewed as an important sub-routine, which although not always optimal, gives a concrete protocol which allows any measurement to reproduce any other measurement. 

\subsection{The general set-up}\label{ss:gen}
In the above we described our set-up in terms of the POVM elements of the trine. This was sufficient for our needs previously, as we in fact did not make use of the post-measurement states of the measurement, but only the result of the measurement. In general, after a system is measured, it will leave the system in a post-measurement state, dependent upon the outcome observed. From a modern perspective, the most general measurement that can be performed is an \emph{instrument} $\mathcal{M}$, which is a collection of completely positive and trace non-increasing maps $\mathcal{M} = \{\Lambda_a\}_a$ (also commonly referred to as \emph{sub-channels}), one associated to each outcome $a$. We will denote by $m$ the number of outcomes of the measurement, which can be arbitrary.  When a state $\rho$ is measured, the probability of the outcome $a$ is
\begin{equation}
	P(a) = \tr[\Lambda_a(\rho)],
\end{equation}
and the state after the measurement is
\begin{equation}
	\rho'_a = \frac{\Lambda_a(\rho)}{P(a)}.
\end{equation}
In order for the measurement to be properly normalised, we must have $\sum_a P(a) = 1$ for all states, which is satisfied as long as $\Lambda = \sum_a \Lambda_a$ is a trace-preserving map. 

Note that since each sub-channel of the instrument is completely positive and trace non-increasing, it can be written as a sum of \emph{Kraus operators},
\begin{equation}
	\Lambda_a(\cdot) = \sum_i K_i^a (\cdot) {K_i^a}^\dagger,
\end{equation}
where $\sum_i {K_i^a}^\dagger K_i^a \leq \mathbb{I}$. The normalisation condition then reads $\sum_{i,a} {K_i^a}^\dagger K_i^a = \mathbb{I}$. This highlights why instruments can be viewed as more general than Kraus operators, since they allow each individual outcome of the measurement to be associated with a collection of Kraus operators, rather than a single Kraus operator.\footnote{This is also sometimes referred to as non-fine-grained measurements, since it can be viewed as a measurement with a pair of outcomes $a$ and $i$, such that $i$ is ignored.} Nevertheless, to each outcome we can still associate a POVM element $M_a$, which determines the probability of a given outcome (but not the post-measurement state). We see that
\begin{equation}
	M_a = \sum_i {K_i^a}^\dagger K_i^a,
\end{equation}
where $P(a) = \tr[M_a \rho]$. 
\begin{figure*}[t!]
	\includegraphics[scale=1]{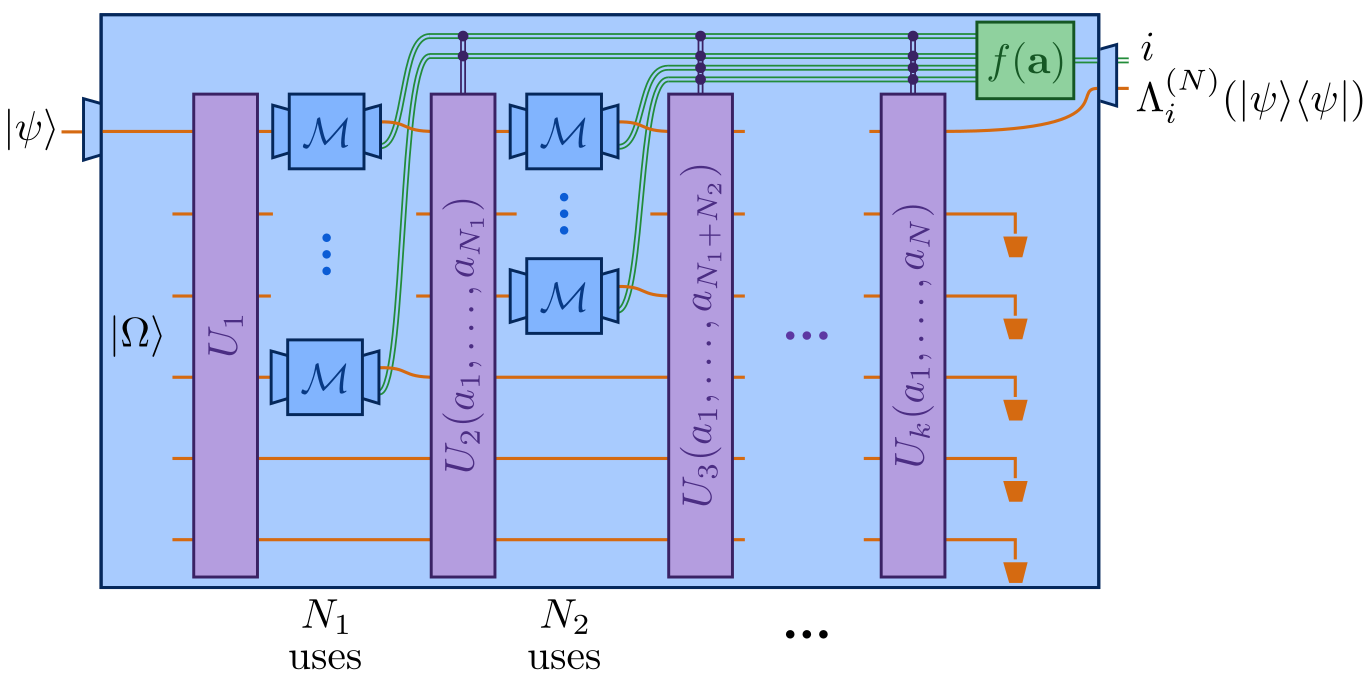}
	\caption{\label{f:protocol general}The general class of protocols we consider here. $N_A$ ancillary particles are prepared in the state $\ket{\Omega}$, and an initial unitary $U_1$ is applied to all $N_A + 1$ particles. In stage one, the first $N_1$ particles are measured, and then a conditional unitary $U_2(a_1,\ldots,a_{N_1})$ is applied. This is iterated $k$ times, such that in stage $j$ the first $N_j$ particles are measured, and then a unitary $U_j(a_1,\ldots, a_{N_1 + \cdots + N_j})$ is applied, which depends upon all previous measurement outcomes. After $N$ measurements have been performed, the final unitary is used to prepare the output particle (taken to be the first without loss of generality), and the measurement outcome is computed according to the function $i = f(\mathbf{a})$. This implements a measurement $\mathcal{M}^{(N)} = \{\Lambda_i^{(N)}\}$.  } 
\end{figure*}
For a general instrument, we find that the post-measurement state $\rho_a'$ will depend both upon the measurement outcome \emph{and} the state measured. This is in contrast to the case of a complete von Neumann measurement, where the post-measurement states depend only upon the measurement outcome (and are independent of the state measured). It is in this sense that we can say that the post-measurement states of the complete von Neumann measurement are `trivial', and we see that a potentially new aspect of the reproduction problem arises when we want to reproduce the post-measurement states of a measurement where these are non-trivial, and retain a dependence upon the measured state. 

The general scenario we wish to consider there is that we want to reproduce a target measurement $\mathcal{T} = \{\Gamma_i\}$, given the ability to perform the measurement $\mathcal{M} = \{\Lambda_a\}$ $N$ times. 

The only requirement we place on $\mathcal{M}$ is that it is \emph{non-trivial}, meaning that its POVM elements are not all proportional to the identity. Such measurements are not measurements at all from a physical perspective, since they don't require any measurement to be performed. They amount simply to producing a random (fictitious) measurement result, and preparing some quantum state. Clearly such trivial measurements will never be able to reproduce any non-trivial measurement, and hence we exclude them from investigations from here on. 

Similarly to before, we consider that we are given a single copy of an unknown pure state $\ket{\psi}$, and at the end of the protocol, in an ideal reproduction we wish to produce an outcome $i$ with probability $\pid(i) = \tr[\Gamma_i(\ket{\psi}\bra{\psi})]$, and to leave the system in the state $\rho'_i = \Gamma_i(\ket{\psi}\bra{\psi})/\pid(i)$. This will not be possible in general, and so we seek to approximate this as closely as possible using an optimal protocol involving $N$ uses of the available measurement. 

\subsection{General protocols}\label{ss:gen protocols}

We now outline the general class of protocols that we consider, as well as the figure of merit we will focus on. The primary difference from before arises due to the fact that we now consider instruments, with post-measurement states. This means that the most general protocols will have a \emph{sequential} aspect to them. In particular, we may consider protocols involving a number of stages, where at one stage we make $N_1$ measurements on a subset of the particles. Depending upon the outcome of these measurements (and previous measurements), we may then apply a unitary on the whole system, before measuring a new subset of $N_2$ particles, and so on. The general structure of protocols we consider, depicted in Fig.~\ref{f:protocol general},  is the following: 
\begin{mdframed}
	\textbf{General Protocol:}
\begin{enumerate}[(i)]
	\item 	Prepare $N_A$ ancillary particles in a state $\ket{\Omega}$, independent of $\ket{\psi}$. Without loss of generality, this can be taken to be $\ket{\Omega} = \ket{0}^{\otimes N_A}$. 
	\item Perform a global unitary $U_1$ on the $N_A + 1$ particles $\ket{\psi}\ket{\Omega}$. 
	\item Measure a subset $\mathcal{S}_1$ of the particles using the available measurement $\mathcal{M}$ $N_1$ times, leading to a string of outcomes $\mathbf{a} = (a_1,\ldots,a_{N_1})$. Due to the freedom in specifying the unitary $U_1$ in the previous step,  the measurements can always be taken to be on the first $N_1$ particles (since systems can be permuted before the measurement). 
	\item Depending upon the string of outcomes $\mathbf{a}$, apply a unitary transformation $U_2(\mathbf{a})$ on all of the particles. 
	\item Measure the first $N_2$ particles using the measurement $\mathcal{M}$. Append the results of these measurements onto the existing string, so that $\mathbf{a} = (a_1,\ldots, a_{N_1},a_{N_1+1},\ldots,a_{N_1 + N_2})$. 
	\item Repeat steps (iv) and (v) until $N = N_1 + N_2 + \cdots + N_k$ measurements have been made, i.e. so that in total there are $k$ measurement stages, and in total $N$ uses of $\mathcal{M}$ are made. 
	\item Dependent on the full string of outcomes $\mathbf{a} = (a_1,\ldots,a_N)$, return as outcome the result $i$. As before, this can be done deterministically, where $i = f(\mathbf{a})$, or probabilistically, according to $p(i|\mathbf{a})$. 
	\item Apply a final unitary transformation $U_{k+1}(\mathbf{a})$, and return as the post-measurement state the first system (without loss of generality). 
\end{enumerate}
\end{mdframed}

This protocol will enact a measurement $\mathcal{M}^{(N)} = \{\Lambda_i^{(N)}\}$ on the system. We can quantify the performance of the protocol using a generalised form of TVD error from before, taking into account now also the post-measurement state, namely
\begin{equation}
	\epstvd{N} =  \int d\psi \sum_i \Big\|\Lambda_i^{(N)}(\ket{\psi}\bra{\psi}) - \Gamma_i(\ket{\psi}\bra{\psi})\Big\|_1,
\end{equation}
where $\|\cdot \|_1$ denotes the trace norm. The above class of protocols are hard to analyse in general, due to their generality. In what follows we will restrict our attention to specific families of protocols which are simpler and less general. Nevertheless, we will see that they are powerful enough to show that any measurement can reproduce arbitrarily well any other measurement. Before doing so, we will first exhibit an important sub-routine, which will reduce the complexity of our problem substantially. 

\subsection{Post-measurement sub-routine}\label{ss:post-meas}
When considering general measurement reproduction, with non-trivial post-measurement states, at first sight it appears that the problem is inherently more difficult than reproducing a complete von Neumann measurement, with its trivial post-measurement state, that depends solely upon the measurement outcome, and not upon the system being measured. In this subsection we will show that there is in fact a \emph{universal sub-routine}, which shows that we can reduce the problem of reproducing a given measurement to the problem of reproducing a perfect complete von Neumann measurements. In particular, if we are given access to a complete von Neumann measurement, the following sub-routine is able to \emph{perfectly} reproduce any other measurement. Thus, in order to use one measurement to reproduce another, we simply only need to show that we can use any measurement to reproduce a complete von Neumann measurement, and then use this sub-routine, in a bootstrap fashion, to reproduce the desired measurement. 

Let us assume that we wish to reproduce a measurement $\mathcal{M}$ as above, and we have the ability to perform a complete von Neumann measurement $\mathbb{V} = \{\ket{i}\bra{i}\}$. Consider the following protocol, which we will refer to as the \emph{post-measurement sub-routine}:
\begin{mdframed}
	\textbf{Post-measurement sub-routine:} 
\begin{enumerate}[(i)]
	\item 	Prepare 2 ancillary particles in the state $\ket{0}\ket{0}$. 
	\item Perform a global unitary $V$ such that
	\begin{equation}\label{e:subroutine V}
		V\ket{\psi}\ket{0}\ket{0} = \sum_{a,j} K_j^a \ket{\psi}\ket{a}\ket{j}.
	\end{equation} 
	\item Ignore the third particle, and measure the second using a perfect complete von Neumann measurement. 
	\item Return as outcome the result of the von Neumann measurement, and return as the post-measurement state the first particle. 
\end{enumerate}
\end{mdframed}
 
It is straightforward to see that the probability of returning the outcome $a$ is 
\begin{equation}
	P(a) = \sum_{j} \tr[K_j^a \ket{\psi}\bra{\psi}{K_j^a}^\dagger] = \tr[M_a \ket{\psi}\bra{\psi}],
\end{equation}
and that the state after measurement is
\begin{equation}
	\rho_a' = \frac{1}{P(a)} \sum_j K_j^a \ket{\psi}\bra{\psi} {K_j^a}^\dagger = \frac{\Lambda_a(\ket{\psi}\bra{\psi})}{P(a)}.
\end{equation}
That is, this shows that a complete von Neumann measurement is able to perfectly reproduce any other measurement. Interestingly,  in this protocol the dimension of the system the von Neumann measurement needs to act upon is determined by the number of outcomes of the target measurement. Note that this does not introduce any difficulties. If for example we can only perform von Neumann measurements of dimension $d$, then in \eqref{e:subroutine V} we would need to encode $\ket{a}$ into an appropriate collection of qudits and perform the required number of complete von Neumann measurements (e.g. if the target measurement had 7 outcomes, and we had access to a qubit von Neumann measurement, then we would encode into 3 qubits, and measure all 3 of them, with only 7 of the 8 possible outcomes ever occurring; equivalently we could take the final outcome to have Kraus operator 0). 
\subsection{Asymptotic reproduction of any measurement by any other measurement}\label{ss:asympt}
We are now in a position to give our first main result: 
\begin{mdframed}
	\textbf{Main result 1:} Given the ability to perform \emph{any non-trivial} measurement, we can reproduce arbitrarily well \emph{any other} (generalised) measurement, by making sufficiently many uses of the first measurement.
\end{mdframed}
In order to prove the main result, we can first note that, given the above post-measurement sub-routine, all that is left to be shown is that any measurement is able to reproduce a perfect von Neumann measurement arbitrarily well in the asymptotic limit. We will now show this, focusing on a simple (albeit non-optimal) protocol, based upon generalised classical cloning. 

We will begin by considering the simplest situation, where we assume that the measurement we have access to is on systems of the same dimension as the target complete von Neumann measurement, which we will denote by $d$. At the end of this section we will explain how to adapt the analysis to take into account the more general situation, where the dimensions differ.

The first key observation is that the ability to perfectly perform a complete von Neumann measurement is equivalent to being able to perfectly distinguish between the $d$ basis states $\{\ket{i}\}$. In the forward direction this is immediate, since a perfect complete von Neumann measurement can indeed distinguish between the $d$ basis states $\{\ket{i}\}$. In the backward direction, a perfect complete von Neumann measurement is the only measurement that is able to perfectly distinguish the basis $\{\ket{i}\}$, and hence the equivalence follows.\footnote{Note that, strictly speaking, we also ignore `trivial' counter-examples to the above, such as the three outcome measurement with outcomes $M_0 = \ket{0}\bra{0}$, $M_1 = \frac{1}{2}\ket{1}\bra{1}$ and $M_2 = \frac{1}{2}\ket{1}\bra{1}$. That is, examples where the measurement is obtained by a trivial post-processing of the von Neumann measurement. In this example, $M_1 = M_2$, and this measurement is equivalent to performing a complete von Neumann measurement, except one outcome involves a fair coin-flip before announcing the result.} 

Thus, we can focus on the task of perfectly distinguishing a basis of states, and if we show that we can do this asymptotically, then we know this implies that we have the ability to perform an ideal measurement.  The following `generalised classical cloning protocol' achieves this goal:

\begin{mdframed}
	\textbf{Generalised classical cloning sub-routine:}
	\begin{enumerate}[(i)]
		\item Prepare $N-1$ ancillary particles in the state $\ket{\Omega} = \ket{0}^{\otimes (N-1)}$. 
		\item Perform the unitary $U$ such that
\begin{equation}
	\ket{\Psi_i} = U \ket{i}\ket{\Omega} = \ket{e_i}^{\otimes N},
\end{equation}
where $\ket{e_i}$ form an orthonormal set of $d$ states (and will be specified more fully in the main text).
\item Measure all $N$ particles using the available measurement $\mathcal{M}$, keeping only the string of measurement results $\mathbf{a} = (a_1,\ldots,a_N)$, and discarding all post-measurement states. 
\item Use maximum-likelihood estimation to guess which basis state $\ket{i}$ was initially prepared.
\end{enumerate}
\end{mdframed}

We will now analyse this sub-routine, to show that in the asymptotic limit of large $N$, it is able to correctly identify the basis state $\ket{i}$ perfectly, independent of the measurement $\mathcal{M}$ used. 

The probability of obtaining the outcome $a$ when measuring the incoming basis state $\ket{i}$ is
\begin{equation}
	P(a|i) = \bra{e_i}M_a\ket{e_i},
\end{equation}
where $\{M_a\}_a$ are the POVM elements of the instrument $\mathcal{M}$. The probabilities of the  string of outcomes $\mathbf{a} = (a_1,\ldots,a_N)$ will be
\begin{equation}
	P^{(N)}(\mathbf{a}|i) = \prod_{k=1}^{N} P(a_k|i).
\end{equation}
Now, the key insight is to view this as $N$ samples from one of $d$ different probability distributions, with each distribution being labelled by $i$. That is, we can map the problem of distinguishing basis states onto the equivalent problem, of having to distinguish between $d$ different classical random variables, given $N$ samples of one of the variables. This is thus an instance of d-ary hypothesis testing. 

Crucially, in the asymptotic limit,  all non-identical random variables can be perfectly distinguished.  Moreover, the probability of making an error drops off exponentially in the number of samples $N$ when using maximum-likelihood estimation \cite{mhypoth}, with the rate given by the minimum pairwise Chernoff distance between the $d$ probability distributions. 

Finally, since the probability of making an error decreases exponentially in the asymptotic limit, so too does the average TVD error $\epstvd{N}$. In particular, we find that $\epsilon_{a|i}$ approaches 0 exponentially fast for $a \neq i$, and similarly $\epsilon_{a|a}$ approach 1 exponentially fast, in both cases as $N$ tends towards infinity. As we show in the appendix, we can find an upper bound on the TVD error that depends only upon $\epsilon_{a|i}$. A direct calculation shows that, given this upper bound approaches 0 exponentially fast, and hence $\epstvd{N}$ must too. 

The only subtlety that arises then is whether the distributions $P(a|i)$ are different for all states $\ket{i}$, which is necessary in order to distinguish the states. In order to understand this issue, let us consider two simple examples. First, returning to the trine measurement from Sec.~\ref{s:trine}, and consider standard classical cloning, with $\ket{e_i} = \ket{i}$, then the probabilities obtained are
\begin{align}
	P(0|0) &= \frac{2}{3},& P(1|0) &= \frac{1}{6},& P(2|0) &= \frac{1}{6}, \nonumber \\
	P(0|1) &= 0,& P(1|1) &= \frac{1}{2},& P(2|1) &= \frac{1}{2}.
\end{align}
Since the two distributions $P(a|0)$ and $P(a|1)$ are distinct, asymptotically the states $\ket{0}$ and $\ket{1}$ can be perfectly distinguished. As a second example, consider the degenerate qutrit measurement with POVM elements
\begin{align}
	M_0 &= \ket{0}\bra{0},& M_1 &= \ket{1}\bra{1} + \ket{2}\bra{2}.
\end{align}
Using standard classical cloning, this measurement is not able to distinguish between the basis states $\ket{1}$ and $\ket{2}$, since they both lead to the same probability distribution $P(0|1) = P(0|2) = 0$ and $P(1|1) = P(1|2) = 1$. However, it is able to distinguish the basis state $\ket{0}$ from $\ket{1}$ and $\ket{2}$. If we were therefore to permute the states $\ket{0}$ and $\ket{1}$ we would be able to distinguish instead $\ket{1}$ from the other two. Using this idea, we can therefore consider a generalised cloning protocol, which consists of performing different hypothesis tests in parallel, by dividing the $N$ clones into $2$ groups. That is, we can consider the following protocol
\begin{align}
    U\ket{0}\ket{0}^{\otimes (N-1)} &= \ket{0}^{\otimes N/2}\ket{1}^{\otimes N/2},\nonumber \\
    U\ket{1}\ket{0}^{\otimes (N-1)} &= \ket{1}^{\otimes N/2}\ket{0}^{\otimes N/2}, \\
    U\ket{2}\ket{0}^{\otimes (N-1)} &= \ket{2}^{\otimes N}.\nonumber
\end{align}
We then use the first $N/2$ measurement results to decide if the state is $\ket{0}$ or not, and the second $N/2$ results to decide if it is $\ket{1}$ or not. At the end of the procedure, we therefore identify the state, with error which still drops exponentially in the asymptotic limit. 

In the Appendix we show more generally that this procedure allows us to identify any basis $\ket{i}$, with error that drops exponentially in the asymptotic limit, using any measurement that isn't trivial. 

Although this protocol will in general definitely not be the optimal one -- in terms of minimising the average TVD error -- we find it nevertheless fascinating that such a simple protocol exists, and can be used as a subroutine to show that any measurement can reproduce any other measurement, with arbitrarily small error, in the asymptotic regime. 

Finally, we must also address the case where the dimension of the measurement we have access to differs from the number of outcomes of the target measurement. If the dimension is larger, then the above analysis goes through, as we can just take the states $\ket{e_i}$ to span a subspace smaller than the dimension of the measurement, without any difficulty. The problematic case arises when the number of outcomes is larger than the dimension of the measurement. In this case, we have to consider a sufficient number of uses of the measurement at a time. As a concrete example, if we want to reproduce a measurement with 5 outcomes, and the measurement we have access to acts on qutrits, then we will consider 2 uses of the measurement, which we treat (conceptually) as a single measurement on a 9 dimensional space, and the states $\ket{e_i}$ forming a 5 dimensional subspace of this 9 dimensional space.\footnote{If this is viewed as problematic per se, then we can consider a modified protocol, where $\ket{\Omega}$ is a state on $(\mathbb{C}^9)^{\otimes N/2}$, and a unitary $U$ which acts on $\mathbb{C}^5 \otimes (\mathbb{C}^9)^{\otimes N/2}$, such that $U\ket{i}\ket{\Omega} = \ket{0}\ket{i}^{\otimes N/2}$. In this way, we don't need to treat the incoming system as a $9$ dimensional system, but simply map its state into a 9-dimensional Hilbert space, and `reset' the incoming particle to a fixed state $\ket{0}$. } The above analysis then applies, without any additional difficulties introduced.

\section{Block-coding and finite-rate protocols}\label{s:block}
Up until now, we have considered the problem of reproducing a single use of a target measurement device, and shown that it can be perfectly reproduced, in the limit where we use the available measuring device an unbounded number of times. If we therefore consider the \emph{rate} of the protocol -- defined as the number of measurements performed in the protocol per measurement reproduced, we see that the protocols we have considered are \emph{zero-rate}. An interesting and important question is whether it is possible to have \emph{finite-rate} measurement reproduction, with protocols which make use of only a finite number of measurements in the protocol per measurement reproduced. 

In this section, we will see that this is possible, if we allow ourselves the analogue of \emph{block-coding} from classical and quantum information theory. In this context, this means that we will not try to reproduce measurements one-by-one, but instead will attempt to reproduce only \emph{multiple measurements in parallel}. In particular, if the target measurement is $\mathcal{T}$, when we will aim to reproduce $\mathcal{T}^{\otimes k}$, that is, the product measurement on $k$ systems, with outcomes $\mathbf{i} = i_1 i_2 \cdots i_k$, and associated sub-channels
	\begin{equation}
		\Gamma_{\mathbf{i}} = \Gamma_{i_1} \otimes \Gamma_{i_2} \otimes \cdots \otimes \Gamma_{i_k}.
	\end{equation}
We then want to consider the question of reproducing this parallel measurement, and are specifically interested in the asymptotic scenario, where $k$ becomes large, and ask for how many measurements $N$ are necessary for a perfect reproduction, and whether a vanishing error can be achieved even if the ratio $\frac{k}{N}$ is kept constant -- i.e.~ such that the number of measurements performed by the protocol is proportional to the number of target measurements. The motivation for studying this limit comes from classical (and quantum) information theory, were we know that by using block-coding schemes, finite rate protocols for communicating over noisy channels are possible.

Our main result in this section is that we can directly make use of finite-rate classical coding schemes in order to show that measurement reproduction can also be achieved at finite rate. In particular, we have
\begin{mdframed}
	\textbf{Main result 2:} Measurement reproduction can be achieved at finite rate when reproducing multiple measurements in parallel. The asymptotic rate $R$ is lower bounded by the classical capacity 
	\begin{equation*}
		C = \max_{p(x)} I(A:X),
	\end{equation*}
	of the classical channel $p(a|x) = \bra{x}M_a \ket{x}$ associated to the measurement, and where $I(A:X)$ is the mutual information of the joint probability distribution $p(a,x) = p(x)p(a|x)$ between the input and output of the channel, $I(A:X) = H(X) - H(X|A)$, with $H(X)$ and $H(X|A)$ the Shannon entropy and conditional entropy, respectively.
\end{mdframed}

In order to prove this result, the first step is to realise that the previous analysis,  which showed that the problem of reproducing an arbitrary measurement reduces to the problem of reproducing just a complete von Neumann measurement, carries over into the present setting. In particular, in order to show that it is possible to find a finite-rate measurement reproduction protocol, we only need to show that there exists a finite-rate protocol for reproducing a complete von Neumann measurement. This follows, since sub-routine 1 only makes use of the von Neumann measurement in order to perform an arbitrary measurement and does not involve any further measurement. It therefore does not alter any conclusion about the rate of the combined protocol. 

As stated in Main result 2, to every measurement $\mathbb{M} = \{M_a\}_a$ (specified only in terms of its POVM elements), we can naturally associate a classical channel -- that is a conditional probability distribution --  $p(a|x)$, via 
\begin{equation}\label{e:ass chan}
	p(a|x) = \bra{x}M_a \ket{x},
\end{equation}
where the input to the channel is $x$, which labels the basis state, the output of the channel is the measurement result $a$. 

From the noisy-channel coding theorem of classical information theory, we know that there exists a sequence of codes (of increasing length), with the error in transmission decreasing with the size of the code, at a rate which approaches the \emph{capacity}, 
\begin{equation}
	C = \max_{p(x)} I(A:X),
\end{equation}
where $I(A:X)$ is the mutual information of the joint probability distribution $p(a,x) = p(x)p(a|x)$ between the input and output of the channel, $I(A:X) = H(X) - H(X|A)$, with $H(X)$ and $H(X|A)$ the Shannon entropy and conditional entropy, respectively. 

Now, the noisy-channel coding theorem guarantees the existence of encodings and decodings, namely functions from $\mathbf{x} =~x_1 \cdots x_k$ (strings of length $k$), to codewords $c_{\mathbf{x}}$ (strings of length $N = R_k k$), and a decoding function $g(\cdot)$, which maps back from length-$N$ strings into length-$k$ strings, with $R_k$, the rate at block-length $k$, approaching the capacity of the channel $C$ as $k$ tends to infinity. 

We can make use of this in the below single-round protocol, making use of $N = R_k k$ imperfect measurements. Before stating the protocol, we note that since we are now aiming to perform $k$ complete von Neumann measurements in parallel, the input will be a state of the form
\begin{equation}
	\ket{\psi} = \sum_{\mathbf{i}} \alpha_{\mathbf{i}} \ket{i_1}\cdots \ket{i_k},
\end{equation} 
where $\alpha_{\mathbf{i}}$ are arbitrary amplitudes. 

\begin{mdframed}
	\textbf{Finite-rate classical-coding-based protocol:}
	\begin{enumerate}[(i)]
		\item Prepare $N-k = (R_k-1)k$ ancillary particles in the state $\ket{\Omega}$.
		\item Apply a unitary $U$ which sends the $N = R_k k$ particles into states encoding the classical codewords,
		\begin{align}
			U\ket{i_1}\cdots\ket{i_k}\ket{\Omega} &= \ket{c_\mathbf{i}},\nonumber \\
			&= \ket{(c_{\mathbf{i}})_1}\cdots \ket{(c_{\mathbf{i}})_N},
		\end{align}
	where $(c_{\mathbf{i}})_j$ denotes the $j^\mathrm{th}$  element of the codeword $c_\mathbf{i}$.
	\item Measure each of the $N$ particles using the available measurement $\mathbb{M}$, producing a string of outcomes $\mathbf{a} = a_1 \cdots a_N$.
	\item Apply the decoding function $g(\cdot)$ of the classical code to the string of outcomes, to produce $\mathbf{j} = g(\mathbf{a})$, a length-$k$ string, which is returned as the outcome of the $k$ measurements.  
	\item Prepare the state $\ket{j_1} \cdots \ket{j_k}$, and return as the post-measurement state. 
	\end{enumerate}
\end{mdframed}

By construction, with error which vanishes as $k$ becomes large, the string $\mathbf{j}$ will be perfectly correlated with the computational basis state $\ket{\mathbf{i}} = \ket{i_1}\cdots\ket{i_k}$, since the unitary $U$ turned this basis state into states encoding the classical codewords. The measurements performed on these codeword states produced measurement outcomes, which can be identified with the outcomes of the associated channel \eqref{e:ass chan}. However, these codewords were precisely the encodings which could then be decoded after the application of the channel, which in this case corresponds to identifying the original basis state. 

Putting everything together, we therefore see that this (approximately) performs a complete von Neumann measurement on $k$ particles, since the probability of having a given computational basis state is precisely $|\alpha_\mathbf{i}|^2$, with error which vanishes as $k$ tends to infinity. Crucially, unlike our previous analysis, the number of additional particles required per particle being measured is now finite, given by the rate $R_k$ of the code used, which we know approaches the capacity $C$ in the limit of large $k$. This shows that we can achieve measurement reproduction at finite rate. 

The above protocol, based upon classical channel coding schemes, in general would not be expected to be optimal. Nevertheless, their existence highlights how classical channel coding theory can be used directly for measurement reproduction, even if not at the optimal possible rate. A highly relevant open question is to understand the optimal rates that can be achieved, and how more general protocols might achieve them.  In particular, we can see immediately that the above protocol is highly constrained in a number of ways. First, it only uses a single round of measurements, second, it does not rely on the post measurement states at all, and lastly the encodings are into product states, not general entangled states. It is reasonable to believe that relaxing all of these could provide significant improvements in the ultimate rates achievable.

It is also worth noting that even a straightforward modification of the above idea can in principle improve the rate. In particular, \eqref{e:ass chan} is only one particular classical channel that we can associate with a quantum measurement. In improved rate can be achieved by optimising over associated classical channels, by varying the states measured, that is by considering the class of channels $\mathcal{C_\mathbb{M}}$,
\begin{equation}
	\mathcal{C}_\mathbb{M} = \{ p(a|x) \, | \, p(a|x) = \tr(M_a \rho_x)\},
\end{equation}
where $\rho_x$ are quantum states for all $x$, and using the associated encoding and decoding scheme. 
\section{Probabilistic protocols}\label{s:prob protocols} 

In this final section we  return to the possibility of using slightly generalised protocols, which \emph{probabilistically} assign outcomes for the target measurement. We will see that, for the average TVD error considered here, such probabilistic protocols can in fact outperform any deterministic protocol.

We will present the main idea by returning to the example from Sec.~\ref{s:trine} which made multiple uses of a trine measurement in order to approximately reproduce a complete von Neumann measurement.  For the case of two uses of the trine, we gave an optimal (deterministic) protocol in \eqref{e:optimal deterministic trine}: That is, we gave an assignment of an outcome $i$, given the  pair of trine measurement outcomes $(a_1, a_2)$.  A  more general possibility is to decide which outcome to assign according to a probabilistic protocol. That is, given the outcomes of the trine measurements, we now need to specify the probabilities $q(i|a_1,a_2)$. From this perspective, we see that the class of protocols considered up until now allowed only for deterministic assignments, such that $q(i|a_1,a_2) = \delta_{i, f(a_1,a_2)}$, for some function $f(a_1,a_2)$. 

We now demonstrate, surprisingly, that probabilistic protocols are more powerful than deterministic ones, by presenting a probabilistic protocol which outperforms the optimal deterministic protocol in \eqref{e:optimal deterministic trine}. The probabilistic protocol we consider is closely related to the deterministic protocol \eqref{e:optimal deterministic trine}. The difference is that now, when we obtain one of the four strings $(a_1,a_2)$ which were previously mapped to $1$ all the time, we now map them to 1 with probability $z$, and flip them to 0 with probability $(1-z)$. The five strings which were previously always mapped to 0 remain so. Mathematically,
\begin{align}
	&q(1|1,1) = q(1|1,2) = q(1|2,1) = q(1|2,2) = z,\\
	&q(0|1,1) = q(0|1,2) = q(0|2,1) = q(0|2,2) = (1-z),\nonumber\\
	&q(0|0,0) = q(0|0,1) = q(0|0,2) = q(0|1,0) = q(0|2,0)= 1.\nonumber
\end{align}
(The previous deterministic protocol corresponds to taking $z=1$.) In terms of POVM elements, this has the effect of changing $Q_0^{(2)}$ and $Q_1^{(2)}$ from \eqref{e:Q2} into
\begin{align}
	\tilde{Q}_0^{(2)} &= zQ_0^{(2)} + (1-z)\mathbb{I},& \tilde{Q}_1^{(2)} &= z Q_1^{(2)}.
\end{align}

A direct calculation allows us to analytically solve for the optimal choice of $z$. We find that 
\begin{equation}
z^*=\frac{9}{8} - \frac{9}{8\sqrt{41}}\approx 0.95,
\end{equation}
is the optimal value for this family of protocols parameterised by $z$, and leads to a slightly improved average TVD error of \begin{equation}
    \langle \tilde{\epsilon}_{(2)}\rangle^*=\frac{\sqrt{41}-5}{16} \approx 0.088,
\end{equation}
which should be compared to the deterministic case ($z=1$) that has error $\epstvd{2}^*=\frac{2(\sqrt{2}-1)}{9} \approx 0.092$. 

Numerical investigation indicates this is not the optimal probabilistic protocol, and even lower average TVD errors can be achieved. Nevertheless, it demonstrates that (simple) probabilistic protocols can provably outperform the best deterministic ones.

One comment is in order. The advantage offered by probabilistic protocols can only arise when the figure of merit considered is non-linear. If the figure of merit would be linear, then a simple argument shows that a deterministic protocol will always perform at least as well as any probabilistic one. Since our figure of merit here is well motivated, quantifying the average discrepancy between the target and implemented measurement, it is genuinely of interest to realise that injecting randomness into the process can be beneficial. We leave it as future research to fully investigate and understand the nature of this probabilistic advantage. 

\section{Conclusions}\label{s:conclusions}
In this paper we have considered the problem of using one measurement to implement another. In particular, we have been interested in the situation where the available measurement is used multiple times, a type of implementation (or simulation) which has received limited attention to date. Surprisingly we have shown that a sufficient number of uses of any measurement allows for the implementation of any other measurement, with error dropping off exponentially fast in the number of measurement uses. As a concise summary, this shows that all measurement are asymptotically equivalent to each other. In order to show this general result, we demonstrated two key sub-routines, which are interesting in their own right - a subroutine for implementing arbitrarily well a complete von Neumann measurement, and a subroutine making use of a complete von Neumann measurement to implement an arbitrary measurement -- including the post measurement state. Furthermore, we showed that measurement implementation can be achieved at finite rate in a block-coding setting -- in order to implement multiple uses of a measurement in parallel, it is sufficient to make use of a finite number of measurements per measurement performed. To show this, we demonstrated how results from classical Shannon theory can be directly harnessed to build measurement implementation protocols. Finally, we showed that allowing protocols to further be probabilistic can be beneficial, by exhibiting a simple probabilistic protocol using the trine which outperforms the best deterministic protocol. 

Our results are important from a number of different perspectives. First, as a special case of our general results, we can consider the task of measurement purification, where an imperfect measurement is used multiple times in order to implement a high quality complete von Neumann measurement. We have shown that even with a few uses, improvements can be made. This may be directly relevant for current quantum technologies -- including noisy intermediate scale (NISQ) devices -- and suggests an alternative approach to directly engineering less noisy measurements -- instead to overcome noise through a measurement implementation protocol instead. This possibility will be relevant in the situation that entangling unitaries can be performed with relatively high fidelity, whereas measurements are comparatively imperfect.  In real applications, of course, neither process will be perfect, and it is an important future question to probe this trade-off. 

Ultimately, we could envisage that measurement implementation becomes a standard technique for improving measurements. Similarly, if a given measurement is hard to engineer, our methods show how a simpler-to-engineer measurement could in principle be used instead, that is, without viewing the available measurement as noisy. This is precisely what is shown in our detailed analysis of the trine measurement -- it can implement the von Neumann measurement, even though it is not a noisy measurement, and even though at first sight it appears to be a very different measurement, one which is itself extremal in the space of measurements. 

From a fundamental perspective, our results raise interesting questions concerning how the act of measurement leads to classical information about a quantum system. Through our protocols, we see how information can be spread out amongst multiple systems, and in the process completely overcome noise or other limitations which arise from specific fixed measurements. Our results can also be viewed from a resource theoretic perspective, whereby it can be viewed as an enlargement of the class of allowed operations to incorporate multiple uses of the available resourceful measurement. From this perspective, it shows that if no cost is associated with making multiple uses, then this trivialises the theory (making all measurements equivalent). However, if only finite numbers of uses are allowed, our results indicate that the theory has a rich structure, allowing for an interesting set of interconversions between measuring devices in general.

There are many fascinating open directions to explore. First, it would be interesting to understand optimal protocols, and the properties they exhibit. We have seen that classical-cloning based protocols are not always optimal (as demonstrated by a detailed analysis of the trine), but sometimes they are. We do not have a full understanding of what determines the properties of an optimal protocol. Furthermore, we considered here the average root-mean-squared error as our figure of merit to optimise. It is important to understand how robust our findings are to other figures of merit, e.g.~when we look at the worst case error, instead of the average case error. 

The protocols described here made use of detailed structure and properties of the available measurement. An important extension is to devise protocols that work for measurements about which we only have partial information. Preliminary investigation shows that some of our protocols indeed work for a range of available measurements (e.g. a single classical cloning protocol will work for a whole range of noisy measurements), however these are only preliminary observations, and a complete understanding of how to optimally make use of partial information is missing. 

In the paradigm here, we placed no restrictions on the types of unitary interactions we are allowed to perform in the protocol. This is a valid choice to make when interested in the fundamental limitations which arise from the available measurements themselves. It would nevertheless be interesting to further place restrictions on the unitary operations allowed, and to understand what effect, if any, this has on the results. For example, we may demand that all the unitary operations should be realisable by a polynomial sized quantum circuit. This would be particularly interesting to study from a practical perspective, where the complexity of the unitary circuit could become a limiting factor in making the protocols realistic and feasible.

In quantum computers, one often only needs to measure a small subset of the qubits at the end of the computation. This raises the possibility that the remaining qubits could be re-initialised, and used as the ancillary qubits in our measurement scheme. A particularly interesting possibility is to make use of some form of block coding scheme. Then we could use all the qubits to form codewords, before measuring them in order to reproduce a more accurate measurement on the small subset of qubits. 

In the case of block coding, we have only shown that finite rate reproduction is possible by relying on results from classical Shannon theory -- in particular from the noisy-channel coding theorem. Our belief is that significantly better rates could be achieved by considering more general protocols, not based upon classical information theory. In particular, we leave it as an open question whether we can make use of quantum codes for classical communication to improve our results, and whether we can characterise the ultimate optimal rate of measurement reproduction.
\medskip

\acknowledgements
We thank Nicolas Brunner, Sandu Popescu, Ieva \v Cepait\.e, Andrew Daley, Gerard Pelegr\'i, Jonathan Pritchard and Chris Corlett  for insightful discussions. PS is a CIFAR Azrieli Global Scholar in the Quantum Information Science Program. PS gratefully acknowledges support from a Royal Society University Research Fellowship (UHQT/NFQI). NL gratefully acknowledges support from the UK Engineering and Physical Sciences Research Council through Grants No. EP/R043957/1, No. EP/S005021/1, and No. EP/T001062/1.

\bibliography{simulating-measurements-bib}

%apsrev4-2.bst 2019-01-14 (MD) hand-edited version of apsrev4-1.bst
%Control: key (0)
%Control: author (8) initials jnrlst
%Control: editor formatted (1) identically to author
%Control: production of article title (0) allowed
%Control: page (0) single
%Control: year (1) truncated
%Control: production of eprint (0) enabled
\begin{thebibliography}{13}%
\makeatletter
\providecommand \@ifxundefined [1]{%
 \@ifx{#1\undefined}
}%
\providecommand \@ifnum [1]{%
 \ifnum #1\expandafter \@firstoftwo
 \else \expandafter \@secondoftwo
 \fi
}%
\providecommand \@ifx [1]{%
 \ifx #1\expandafter \@firstoftwo
 \else \expandafter \@secondoftwo
 \fi
}%
\providecommand \natexlab [1]{#1}%
\providecommand \enquote  [1]{``#1''}%
\providecommand \bibnamefont  [1]{#1}%
\providecommand \bibfnamefont [1]{#1}%
\providecommand \citenamefont [1]{#1}%
\providecommand \href@noop [0]{\@secondoftwo}%
\providecommand \href [0]{\begingroup \@sanitize@url \@href}%
\providecommand \@href[1]{\@@startlink{#1}\@@href}%
\providecommand \@@href[1]{\endgroup#1\@@endlink}%
\providecommand \@sanitize@url [0]{\catcode `\\12\catcode `\$12\catcode
  `\&12\catcode `\#12\catcode `\^12\catcode `\_12\catcode `\%12\relax}%
\providecommand \@@startlink[1]{}%
\providecommand \@@endlink[0]{}%
\providecommand \url  [0]{\begingroup\@sanitize@url \@url }%
\providecommand \@url [1]{\endgroup\@href {#1}{\urlprefix }}%
\providecommand \urlprefix  [0]{URL }%
\providecommand \Eprint [0]{\href }%
\providecommand \doibase [0]{https://doi.org/}%
\providecommand \selectlanguage [0]{\@gobble}%
\providecommand \bibinfo  [0]{\@secondoftwo}%
\providecommand \bibfield  [0]{\@secondoftwo}%
\providecommand \translation [1]{[#1]}%
\providecommand \BibitemOpen [0]{}%
\providecommand \bibitemStop [0]{}%
\providecommand \bibitemNoStop [0]{.\EOS\space}%
\providecommand \EOS [0]{\spacefactor3000\relax}%
\providecommand \BibitemShut  [1]{\csname bibitem#1\endcsname}%
\let\auto@bib@innerbib\@empty
%</preamble>
\bibitem [{\citenamefont {Yuen}(1987)}]{yuenDesignTransparentOptical1987}%
  \BibitemOpen
  \bibfield  {author} {\bibinfo {author} {\bibfnamefont {H.~P.}\ \bibnamefont
  {Yuen}},\ }\bibfield  {title} {\bibinfo {title} {Design of transparent
  optical networks by using novel quantum amplifiers and sources},\ }\href
  {https://doi.org/10.1364/OL.12.000789} {\bibfield  {journal} {\bibinfo
  {journal} {Opt. Lett., OL}\ }\textbf {\bibinfo {volume} {12}},\ \bibinfo
  {pages} {789} (\bibinfo {year} {1987})}\BibitemShut {NoStop}%
\bibitem [{\citenamefont {Dall'Arno}\ \emph {et~al.}(2010)\citenamefont
  {Dall'Arno}, \citenamefont {D'Ariano},\ and\ \citenamefont
  {Sacchi}}]{dallarnoPurificationNoisyQuantum2010}%
  \BibitemOpen
  \bibfield  {author} {\bibinfo {author} {\bibfnamefont {M.}~\bibnamefont
  {Dall'Arno}}, \bibinfo {author} {\bibfnamefont {G.~M.}\ \bibnamefont
  {D'Ariano}},\ and\ \bibinfo {author} {\bibfnamefont {M.~F.}\ \bibnamefont
  {Sacchi}},\ }\bibfield  {title} {\bibinfo {title} {Purification of noisy
  quantum measurements},\ }\href {https://doi.org/10.1103/PhysRevA.82.042315}
  {\bibfield  {journal} {\bibinfo  {journal} {Phys. Rev. A}\ }\textbf {\bibinfo
  {volume} {82}},\ \bibinfo {pages} {042315} (\bibinfo {year}
  {2010})}\BibitemShut {NoStop}%
\bibitem [{\citenamefont {Abbas}\ \emph {et~al.}(2020)\citenamefont {Abbas},
  \citenamefont {Andersson}, \citenamefont {Asfaw}, \citenamefont {Corcoles},
  \citenamefont {Bello}, \citenamefont {Ben-Haim}, \citenamefont {Bozzo-Rey},
  \citenamefont {Bravyi}, \citenamefont {Bronn}, \citenamefont {Capelluto},
  \citenamefont {Vazquez}, \citenamefont {Ceroni}, \citenamefont {Chen},
  \citenamefont {Frisch}, \citenamefont {Gambetta}, \citenamefont {Garion},
  \citenamefont {Gil}, \citenamefont {Gonzalez}, \citenamefont {Harkins},
  \citenamefont {Imamichi}, \citenamefont {Jayasinha}, \citenamefont {Kang},
  \citenamefont {h.~Karamlou}, \citenamefont {Loredo}, \citenamefont {McKay},
  \citenamefont {Maldonado}, \citenamefont {Macaluso}, \citenamefont
  {Mezzacapo}, \citenamefont {Minev}, \citenamefont {Movassagh}, \citenamefont
  {Nannicini}, \citenamefont {Nation}, \citenamefont {Phan}, \citenamefont
  {Pistoia}, \citenamefont {Rattew}, \citenamefont {Schaefer}, \citenamefont
  {Shabani}, \citenamefont {Smolin}, \citenamefont {Stenger}, \citenamefont
  {Temme}, \citenamefont {Tod}, \citenamefont {Wanzambi}, \citenamefont
  {Wood},\ and\ \citenamefont {Wootton.}}]{Qiskit-Textbook}%
  \BibitemOpen
  \bibfield  {author} {\bibinfo {author} {\bibfnamefont {A.}~\bibnamefont
  {Abbas}}, \bibinfo {author} {\bibfnamefont {S.}~\bibnamefont {Andersson}},
  \bibinfo {author} {\bibfnamefont {A.}~\bibnamefont {Asfaw}}, \bibinfo
  {author} {\bibfnamefont {A.}~\bibnamefont {Corcoles}}, \bibinfo {author}
  {\bibfnamefont {L.}~\bibnamefont {Bello}}, \bibinfo {author} {\bibfnamefont
  {Y.}~\bibnamefont {Ben-Haim}}, \bibinfo {author} {\bibfnamefont
  {M.}~\bibnamefont {Bozzo-Rey}}, \bibinfo {author} {\bibfnamefont
  {S.}~\bibnamefont {Bravyi}}, \bibinfo {author} {\bibfnamefont
  {N.}~\bibnamefont {Bronn}}, \bibinfo {author} {\bibfnamefont
  {L.}~\bibnamefont {Capelluto}}, \bibinfo {author} {\bibfnamefont {A.~C.}\
  \bibnamefont {Vazquez}}, \bibinfo {author} {\bibfnamefont {J.}~\bibnamefont
  {Ceroni}}, \bibinfo {author} {\bibfnamefont {R.}~\bibnamefont {Chen}},
  \bibinfo {author} {\bibfnamefont {A.}~\bibnamefont {Frisch}}, \bibinfo
  {author} {\bibfnamefont {J.}~\bibnamefont {Gambetta}}, \bibinfo {author}
  {\bibfnamefont {S.}~\bibnamefont {Garion}}, \bibinfo {author} {\bibfnamefont
  {L.}~\bibnamefont {Gil}}, \bibinfo {author} {\bibfnamefont {S.~D. L.~P.}\
  \bibnamefont {Gonzalez}}, \bibinfo {author} {\bibfnamefont {F.}~\bibnamefont
  {Harkins}}, \bibinfo {author} {\bibfnamefont {T.}~\bibnamefont {Imamichi}},
  \bibinfo {author} {\bibfnamefont {P.}~\bibnamefont {Jayasinha}}, \bibinfo
  {author} {\bibfnamefont {H.}~\bibnamefont {Kang}}, \bibinfo {author}
  {\bibfnamefont {A.}~\bibnamefont {h.~Karamlou}}, \bibinfo {author}
  {\bibfnamefont {R.}~\bibnamefont {Loredo}}, \bibinfo {author} {\bibfnamefont
  {D.}~\bibnamefont {McKay}}, \bibinfo {author} {\bibfnamefont
  {A.}~\bibnamefont {Maldonado}}, \bibinfo {author} {\bibfnamefont
  {A.}~\bibnamefont {Macaluso}}, \bibinfo {author} {\bibfnamefont
  {A.}~\bibnamefont {Mezzacapo}}, \bibinfo {author} {\bibfnamefont
  {Z.}~\bibnamefont {Minev}}, \bibinfo {author} {\bibfnamefont
  {R.}~\bibnamefont {Movassagh}}, \bibinfo {author} {\bibfnamefont
  {G.}~\bibnamefont {Nannicini}}, \bibinfo {author} {\bibfnamefont
  {P.}~\bibnamefont {Nation}}, \bibinfo {author} {\bibfnamefont
  {A.}~\bibnamefont {Phan}}, \bibinfo {author} {\bibfnamefont {M.}~\bibnamefont
  {Pistoia}}, \bibinfo {author} {\bibfnamefont {A.}~\bibnamefont {Rattew}},
  \bibinfo {author} {\bibfnamefont {J.}~\bibnamefont {Schaefer}}, \bibinfo
  {author} {\bibfnamefont {J.}~\bibnamefont {Shabani}}, \bibinfo {author}
  {\bibfnamefont {J.}~\bibnamefont {Smolin}}, \bibinfo {author} {\bibfnamefont
  {J.}~\bibnamefont {Stenger}}, \bibinfo {author} {\bibfnamefont
  {K.}~\bibnamefont {Temme}}, \bibinfo {author} {\bibfnamefont
  {M.}~\bibnamefont {Tod}}, \bibinfo {author} {\bibfnamefont {E.}~\bibnamefont
  {Wanzambi}}, \bibinfo {author} {\bibfnamefont {S.}~\bibnamefont {Wood}},\
  and\ \bibinfo {author} {\bibfnamefont {J.}~\bibnamefont {Wootton.}},\ }\href
  {http://community.qiskit.org/textbook} {\bibinfo {title} {Learn quantum
  computation using qiskit}} (\bibinfo {year} {2020})\BibitemShut {NoStop}%
\bibitem [{\citenamefont {Chitambar}\ and\ \citenamefont
  {Gour}(2019)}]{chitambarQuantumResourceTheories2019}%
  \BibitemOpen
  \bibfield  {author} {\bibinfo {author} {\bibfnamefont {E.}~\bibnamefont
  {Chitambar}}\ and\ \bibinfo {author} {\bibfnamefont {G.}~\bibnamefont
  {Gour}},\ }\bibfield  {title} {\bibinfo {title} {Quantum resource theories},\
  }\href {https://doi.org/10.1103/RevModPhys.91.025001} {\bibfield  {journal}
  {\bibinfo  {journal} {Rev. Mod. Phys.}\ }\textbf {\bibinfo {volume} {91}},\
  \bibinfo {pages} {025001} (\bibinfo {year} {2019})}\BibitemShut {NoStop}%
\bibitem [{\citenamefont {Skrzypczyk}\ and\ \citenamefont
  {Linden}(2019)}]{skrzypczykRobustnessMeasurementDiscrimination2019}%
  \BibitemOpen
  \bibfield  {author} {\bibinfo {author} {\bibfnamefont {P.}~\bibnamefont
  {Skrzypczyk}}\ and\ \bibinfo {author} {\bibfnamefont {N.}~\bibnamefont
  {Linden}},\ }\bibfield  {title} {\bibinfo {title} {Robustness of
  {{Measurement}}, {{Discrimination Games}}, and {{Accessible Information}}},\
  }\href {https://doi.org/10.1103/PhysRevLett.122.140403} {\bibfield  {journal}
  {\bibinfo  {journal} {Phys. Rev. Lett.}\ }\textbf {\bibinfo {volume} {122}},\
  \bibinfo {pages} {140403} (\bibinfo {year} {2019})}\BibitemShut {NoStop}%
\bibitem [{\citenamefont {Oszmaniec}\ and\ \citenamefont
  {Biswas}(2019)}]{oszmaniecOperationalRelevanceResource2019a}%
  \BibitemOpen
  \bibfield  {author} {\bibinfo {author} {\bibfnamefont {M.}~\bibnamefont
  {Oszmaniec}}\ and\ \bibinfo {author} {\bibfnamefont {T.}~\bibnamefont
  {Biswas}},\ }\bibfield  {title} {\bibinfo {title} {Operational relevance of
  resource theories of quantum measurements},\ }\href
  {https://doi.org/10.22331/q-2019-04-26-133} {\bibfield  {journal} {\bibinfo
  {journal} {Quantum}\ }\textbf {\bibinfo {volume} {3}},\ \bibinfo {pages}
  {133} (\bibinfo {year} {2019})}\BibitemShut {NoStop}%
\bibitem [{\citenamefont {Takagi}\ and\ \citenamefont
  {Regula}(2019)}]{takagiGeneralResourceTheories2019a}%
  \BibitemOpen
  \bibfield  {author} {\bibinfo {author} {\bibfnamefont {R.}~\bibnamefont
  {Takagi}}\ and\ \bibinfo {author} {\bibfnamefont {B.}~\bibnamefont
  {Regula}},\ }\bibfield  {title} {\bibinfo {title} {General {{Resource
  Theories}} in {{Quantum Mechanics}} and {{Beyond}}: {{Operational
  Characterization}} via {{Discrimination Tasks}}},\ }\href
  {https://doi.org/10.1103/PhysRevX.9.031053} {\bibfield  {journal} {\bibinfo
  {journal} {Phys. Rev. X}\ }\textbf {\bibinfo {volume} {9}},\ \bibinfo {pages}
  {031053} (\bibinfo {year} {2019})}\BibitemShut {NoStop}%
\bibitem [{\citenamefont {Guff}\ \emph {et~al.}(2019)\citenamefont {Guff},
  \citenamefont {McMahon}, \citenamefont {Sanders},\ and\ \citenamefont
  {Gilchrist}}]{guffResourceTheoryQuantum2019a}%
  \BibitemOpen
  \bibfield  {author} {\bibinfo {author} {\bibfnamefont {T.}~\bibnamefont
  {Guff}}, \bibinfo {author} {\bibfnamefont {N.~A.}\ \bibnamefont {McMahon}},
  \bibinfo {author} {\bibfnamefont {Y.~R.}\ \bibnamefont {Sanders}},\ and\
  \bibinfo {author} {\bibfnamefont {A.}~\bibnamefont {Gilchrist}},\ }\bibfield
  {title} {\bibinfo {title} {A {{Resource Theory}} of {{Quantum
  Measurements}}},\ }\href@noop {} {\bibfield  {journal} {\bibinfo  {journal}
  {arXiv:1902.08490 [quant-ph]}\ } (\bibinfo {year} {2019})},\ \Eprint
  {https://arxiv.org/abs/1902.08490} {arXiv:1902.08490 [quant-ph]} \BibitemShut
  {NoStop}%
\bibitem [{\citenamefont {Bennett}\ \emph {et~al.}(1993)\citenamefont
  {Bennett}, \citenamefont {Brassard}, \citenamefont {Cr{\'e}peau},
  \citenamefont {Jozsa}, \citenamefont {Peres},\ and\ \citenamefont
  {Wootters}}]{bennettTeleportingUnknownQuantum1993}%
  \BibitemOpen
  \bibfield  {author} {\bibinfo {author} {\bibfnamefont {C.~H.}\ \bibnamefont
  {Bennett}}, \bibinfo {author} {\bibfnamefont {G.}~\bibnamefont {Brassard}},
  \bibinfo {author} {\bibfnamefont {C.}~\bibnamefont {Cr{\'e}peau}}, \bibinfo
  {author} {\bibfnamefont {R.}~\bibnamefont {Jozsa}}, \bibinfo {author}
  {\bibfnamefont {A.}~\bibnamefont {Peres}},\ and\ \bibinfo {author}
  {\bibfnamefont {W.~K.}\ \bibnamefont {Wootters}},\ }\bibfield  {title}
  {\bibinfo {title} {Teleporting an unknown quantum state via dual classical
  and {{Einstein}}-{{Podolsky}}-{{Rosen}} channels},\ }\href
  {https://doi.org/10.1103/PhysRevLett.70.1895} {\bibfield  {journal} {\bibinfo
   {journal} {Phys. Rev. Lett.}\ }\textbf {\bibinfo {volume} {70}},\ \bibinfo
  {pages} {1895} (\bibinfo {year} {1993})}\BibitemShut {NoStop}%
\bibitem [{\citenamefont {Guerini}\ \emph {et~al.}(2017)\citenamefont
  {Guerini}, \citenamefont {Bavaresco}, \citenamefont {Terra~Cunha},\ and\
  \citenamefont {Ac{\'i}n}}]{gueriniOperationalFrameworkQuantum2017}%
  \BibitemOpen
  \bibfield  {author} {\bibinfo {author} {\bibfnamefont {L.}~\bibnamefont
  {Guerini}}, \bibinfo {author} {\bibfnamefont {J.}~\bibnamefont {Bavaresco}},
  \bibinfo {author} {\bibfnamefont {M.}~\bibnamefont {Terra~Cunha}},\ and\
  \bibinfo {author} {\bibfnamefont {A.}~\bibnamefont {Ac{\'i}n}},\ }\bibfield
  {title} {\bibinfo {title} {Operational framework for quantum measurement
  simulability},\ }\href {https://doi.org/10.1063/1.4994303} {\bibfield
  {journal} {\bibinfo  {journal} {Journal of Mathematical Physics}\ }\textbf
  {\bibinfo {volume} {58}},\ \bibinfo {pages} {092102} (\bibinfo {year}
  {2017})}\BibitemShut {NoStop}%
\bibitem [{\citenamefont {Oszmaniec}\ \emph {et~al.}(2017)\citenamefont
  {Oszmaniec}, \citenamefont {Guerini}, \citenamefont {Wittek},\ and\
  \citenamefont
  {Ac{\'i}n}}]{oszmaniecSimulatingPositiveOperatorValuedMeasures2017}%
  \BibitemOpen
  \bibfield  {author} {\bibinfo {author} {\bibfnamefont {M.}~\bibnamefont
  {Oszmaniec}}, \bibinfo {author} {\bibfnamefont {L.}~\bibnamefont {Guerini}},
  \bibinfo {author} {\bibfnamefont {P.}~\bibnamefont {Wittek}},\ and\ \bibinfo
  {author} {\bibfnamefont {A.}~\bibnamefont {Ac{\'i}n}},\ }\bibfield  {title}
  {\bibinfo {title} {Simulating {{Positive}}-{{Operator}}-{{Valued Measures}}
  with {{Projective Measurements}}},\ }\href
  {https://doi.org/10.1103/PhysRevLett.119.190501} {\bibfield  {journal}
  {\bibinfo  {journal} {Phys. Rev. Lett.}\ }\textbf {\bibinfo {volume} {119}},\
  \bibinfo {pages} {190501} (\bibinfo {year} {2017})}\BibitemShut {NoStop}%
\bibitem [{\citenamefont {Leang}\ and\ \citenamefont
  {Johnson}(1997)}]{mhypoth}%
  \BibitemOpen
  \bibfield  {author} {\bibinfo {author} {\bibfnamefont {C.}~\bibnamefont
  {Leang}}\ and\ \bibinfo {author} {\bibfnamefont {D.}~\bibnamefont
  {Johnson}},\ }\bibfield  {title} {\bibinfo {title} {On the asymptotics of
  m-hypothesis bayesian detection},\ }\href {https://doi.org/10.1109/18.567705}
  {\bibfield  {journal} {\bibinfo  {journal} {IEEE Transactions on Information
  Theory}\ }\textbf {\bibinfo {volume} {43}},\ \bibinfo {pages} {280} (\bibinfo
  {year} {1997})}\BibitemShut {NoStop}%
\bibitem [{\citenamefont {Maciejewski}\ \emph {et~al.}(2023)\citenamefont
  {Maciejewski}, \citenamefont {Puchała},\ and\ \citenamefont
  {Oszmaniec}}]{maciejewski2023exploring}%
  \BibitemOpen
  \bibfield  {author} {\bibinfo {author} {\bibfnamefont {F.~B.}\ \bibnamefont
  {Maciejewski}}, \bibinfo {author} {\bibfnamefont {Z.}~\bibnamefont
  {Puchała}},\ and\ \bibinfo {author} {\bibfnamefont {M.}~\bibnamefont
  {Oszmaniec}},\ }\href@noop {} {\bibinfo {title} {Exploring quantum
  average-case distances: proofs, properties, and examples}} (\bibinfo {year}
  {2023}),\ \Eprint {https://arxiv.org/abs/2112.14284} {arXiv:2112.14284
  [quant-ph]} \BibitemShut {NoStop}%
\end{thebibliography}%
\onecolumngrid
\newpage
\section*{APPENDICES}

\appendix

\section{Evaluating or bounding the average total variation distance (TVD) error}
In this appendix we will show how the integral over all states can be performed for qubits, in order to obtain an explicit optimisation problem for the TVD error. 

For higher dimensional system, while this average doesn't appear to be easily computable, we show how it can be upper bounded, using ideas from \cite{maciejewski2023exploring}. This upper bound can then be evaluated, and proves useful for bounding the error of protocols in general situations.

\subsection{Two outcome qubit measurements}
Our starting point is to express \eqref{e:eps N simple} more explicitly. In particular, we will write $\ket{\psi} = \cos \frac{\theta}{2} \ket{0} + e^{i \phi} \sin \frac{\theta}{2}\ket{1}$ (i.e. in Bloch sphere notation), such that $d\psi = \frac{1}{4\pi} \sin \theta d\theta d \phi$.

We will assume that $M_i^{(N)}$ is diagonal in the $\{\ket{0},\ket{1}\}$ basis, with $M_0^{(N)}$ explicit given as 
\begin{equation}\label{e:M0N form}
	M_0^{(N)} = (1-\epsilon_0) \ket{0}\bra{0} + \epsilon_1 \ket{1}\bra{1}.
\end{equation}
Two comments are in order here. First, given any protocol that leads to a measurement with non-vanishing off-diagonal elements, we can always find another protocol with vanishing off-diagonal elements. This just says that we can always dephase any protocol. One way to see this is that we can always begin the protocol by dephasing the state to be measured, by entangling it with another ancillary system. If we instead include this dephasing of the state in the measurement (which essentially amounts to consider the Heisenberg picture of this process), it will have the effect of dephasing the POVM elements. Hence, there is no difficulty in obtaining diagonal POVMs.

On the other hand, there \emph{is} a lack of generality in the class of protocols that we consider by restricting to protocols which dephase the incoming state, and one may wonder if it could in fact be optimal to consider such non-diagonal protocols. We do not have a proof that this cannot happen, but we do note that this would be a rather surprising result if it were true, and would be attributable to our figure of merit. This is because if we were to instead consider a trace-distance based measure, which would focus on the \emph{worst-case} behaviour of the protocol, then it follows straight away that an optimal protocol cannot necessarily have off-diagonal terms. This is because we can view the error instead from the perspective of distinguishability; the fact that there is an error, says that the target measurement and the simulated measurement can be distinguished. We know that under the trace-distance, two measurements cannot become more distinguishable under pre-processing (which is a statement of the data processing inequality). However, dephasing is a pre-processing, which leaves the target measurement unaltered. 

This argument shows that as far as the worst case error is concerned, we can restrict to diagonal measurements without loss of generality. Furthermore, in \ref{a:upperbound} below, we will show how to bound the average TVD error from above. This bound was also shown to satisfy the data processing inequality in \cite{maciejewski2023exploring}. Hence, it seems a reasonable assumption that the average TVD error should have this behaviour too, and we will therefore restrict our attention to diagonal measurements from here on. 

With this in place, we now have
\begin{equation}
	P^{(N)}(0) = (1-\epsilon_0)\cos^2\frac{\theta}{2} + \epsilon_1 \sin^2\frac{\theta}{2},
\end{equation}
and so
\begin{align}
	\epstvd{N} &= \frac{1}{2\pi}\int_0^{2\pi}d\phi \int_0^\pi \sin \theta d\theta \big|P^{(N)}(0)-\pid(0)\big|, \nonumber \\
	&= \int_0^\pi \sin \theta d\theta \left|(1-\epsilon_0)\cos^2\frac{\theta}{2} + \epsilon_1 \sin^2\frac{\theta}{2}-\cos^2\frac{\theta}{2}\right|, \nonumber \\
	&= \int_0^\pi \sin \theta d\theta \left|\epsilon_1 \sin^2\frac{\theta}{2}-\epsilon_0\cos^2\frac{\theta}{2}\right|.
\end{align}
The term $\epsilon_1 \sin^2\frac{\theta}{2}-\epsilon_0\cos^2\frac{\theta}{2}$ changes sign at $\theta = \theta_c$, where $\theta_c$ satisfies
\begin{equation}
	\tan^2\frac{\theta_c}{2} = \frac{\epsilon_0}{\epsilon_1}.
\end{equation} 
We can therefore express the integral as
\begin{equation}
	\epstvd{N} = \int_{\theta_c}^\pi \sin \theta d\theta \left(\epsilon_1 \sin^2\frac{\theta}{2}-\epsilon_0\cos^2\frac{\theta}{2}\right) - \int_0^{\theta_c} \sin \theta d\theta \left(\epsilon_1 \sin^2\frac{\theta}{2}-\epsilon_0\cos^2\frac{\theta}{2}\right).
\end{equation}
This can be now be readily integrated. We find
\begin{equation}
	\epstvd{N} = \frac{1}{4}(\epsilon_0 + \epsilon_1)(1 - 4 \cos \theta_c + \cos 2\theta_c) + 2 \epsilon_1 \cos \theta_c.
\end{equation}
Substituting for $\theta_c$, and simplifying, we finally arrive at 
\begin{equation}
	\epstvd{N} = \frac{\epsilon_0^2 + \epsilon_1^2}{\epsilon_0 + \epsilon_1}.
\end{equation}

\subsection{General upper bound}\label{a:upperbound}
Going beyond qubits, it appears challenging to explicitly evaluate the average TVD error. Nevertheless, we can make progress by finding an upper bound, based upon the results in \cite{maciejewski2023exploring}. For completeness, we reproduce the required results here.  

The quantity of interest in the general case of a qudit von Neumann measurement is 
\begin{equation}
	\epstvd{N} = \sum_i \int d\psi \left| P^{(N)}(i|\psi) - \pid(i|\psi) \right|.
\end{equation}
where the average is over all pure states $\ket{\psi}$ in $\mathbb{C}^d$, and we explicitly exhibit to the fact that $P^{(N)}(i|\psi) = \bra{\psi}M^{(N)}_i\ket{\psi}$ and $\pid(i|\psi) = |\langle i | \psi \rangle|^2$ depend upon the state $\ket{\psi}$. 

We start by making use of Jensen's inequality, which states that for a concave function $f$, and a random variable $X$, $\langle f(X)\rangle \leq f\left(\langle X \rangle \right)$. In particular, for each $i$ we define a random variable $X_i = (P^{(N)}(i|\psi) - \pid(i|\psi))^2$ (with the average over $\psi$), and take $f(x) = \sqrt{x}$, such that $f(x^2) = |x|$. Applying Jensen's inequality, we therefore see that
\begin{align}
	\epstvd{N} = \sum_i \int d\psi \left| P^{(N)}(i|\psi) - \pid(i|\psi) \right| &\leq \sum_i \int d\psi (P^{(N)}(i|\psi) - \pid(i|\psi))^2.\nonumber \\
	&= \sum_i \int d\psi \big[\bra{\psi}M_i^{(N)}\ket{\psi} -\bra{\psi}\Pi_i\ket{\psi}\big]^2.
\end{align}

We can re-express this by noting that
\begin{equation}
	\big[\bra{\psi}M_i^{(N)}\ket{\psi} -\bra{\psi}\Pi_i\ket{\psi}\big]^2 = \tr\left[\ket{\psi}\bra{\psi}^{\otimes 2}\left(M_i^{(N)}-\Pi_i\right)^{\otimes 2}\right]
\end{equation}
The benefit of this is that we can now use the fact that 
\begin{equation}
\int d\psi \ket{\psi}\bra{\psi}^{\otimes 2} = \frac{1}{d_{\mathrm{sym}}}\Pi_{\mathrm{sym}},
\end{equation}
where $d_{\mathrm{sym}} = \frac{d(d+1)}{2}$ is the dimension of the symmetric subspace of two qudits. That is, the integral is proportional to the projector onto the symmetric subspace. Finally, we can use the fact that $\Pi_{\mathrm{sym}} = \frac{1}{2}(\mathbb{I} + S)$, with $S$ the \textsc{SWAP} operator. Putting everything together, after a lengthy but straightforward calculation, we arrive at the bound
\begin{equation}\label{e:epsN qudit}
	\epstvd{N} \leq \frac{1}{2 d_\mathrm{sym}}\sum_i \left[\left(\sum_{j\neq i} \epsilon_{j|i} - \sum_{k\neq i} \epsilon_{i|k}\right)^2 + \left(\sum_{j\neq i} \epsilon_{j|i}\right)^2 + \sum_{k\neq i}\epsilon_{i|k}^2\right],
\end{equation}
where, as above, we assume that $M_i^{(N)}$ is diagonal in the basis of the von Neumann measurement $\{\ket{i}\}$, and can be expressed as 
\begin{equation}
	M_i^{(N)} = \left(1-\sum_{j\neq i} \epsilon_{j|i}\right)\ket{i}\bra{i} + \sum_{k\neq i} \epsilon_{i|k}\ket{k}\bra{k},
\end{equation}
which is in terms of the \emph{errors} $\epsilon_{i|j}$ to return the outcome $i$ when the incoming basis state was $j$. This is a convenient way to represent each POVM element, and focuses on the errors that it makes. Notably, due to the normalisation of measurements, the probability of correctly identifying the state $\ket{i}$ is completely determined by the probabilities of incorrectly announcing the outcome $i$ when the state is any other state $\ket{j} \neq \ket{i}$. 

In \eqref{e:epsN qudit} we can see immediate that if all the individual errors vanish, then on average there is no error in the measurement, as must be the case. Furthermore, we see that if the individual errors (on basis states) can be shown to decrease exponentially with $N$, then the average error must also decrease exponentially with $N$.

\section{Solving analytically for the minimal average TVD error}
In this appendix we will show that it is possible to analytically solve for the minimal average TVD error that can be achieved in the case of approximately implementing a two outcome measurement (including, as a special case, a qubit von Neumann measurement). The method we will apply is a geometrical one, which shows that there are three regimes, which we can explicitly identify.

Let us begin analysing the constraints in \eqref{quadratic}. Geometrically, we see that we are optimising over a rectangle in the $\epsilon_0-\epsilon_1$ plane. Denoting the coordinate by $(\epsilon_0,\epsilon_1)$,  the vertices of this rectangle are seen to be at $(1-\lambda_{\textrm{max}}, \lambda_{\textrm{min}})$, $(1-\lambda_{\textrm{max}}, \lambda_{\textrm{max}})$, $(1-\lambda_{\textrm{min}}, \lambda_{\textrm{min}})$ and $(1-\lambda_{\textrm{min}}, \lambda_{\textrm{max}})$. 

We can also straightforwardly identify the contours of the objective function. In particular, points which satisfy 
\begin{equation}
	\epsilon_0^2 + \epsilon_1^2 = c(\epsilon_0+\epsilon_1),
\end{equation}
are those where the function takes value $c$. This can be re-expressed as
\begin{equation}
	\left(\epsilon_0-\tfrac{1}{2}c\right)^2 + \left(\epsilon_1-\tfrac{1}{2}c\right)^2 = \left(\frac{c}{\sqrt{2}}\right)^2, 
\end{equation}
that is, the contours are circles of radius $\frac{c}{\sqrt{2}}$, centred at $(\epsilon_0^c,\epsilon_1^c) = (\tfrac{1}{2}c, \tfrac{1}{2}c)$. Clearly, $c = 0 = \epstvd{N}^*$ corresponds to $\epsilon_0 = \epsilon_1 = 0$, and this will only be in the rectangle if  $\lambda_{\rm min} = 0$ and $\lambda_{\rm max} = 1$. From \eqref{e:M0N form}, this implies that $M_0^{(N)} = \ket{0}\bra{0}$, as expected.

When  $\lambda_{\textrm{min}} > 0$ and/or $\lambda_{\textrm{max}} < 1$, then there will be a range of $c$ such that the centre of the circle is outside the rectangle. Our optimisation problem can be understood geometrically as finding the size of the smallest circle which \emph{just touches} the rectangle. The point at which the circle and rectangle touch will be the optimal solution $(\epsilon_0^*, \epsilon_1^*)$. 

We can now make two key observations. First, since the centre of the circle always moves from the point $(0,0)$ along the line $\epsilon_0 = \epsilon_1$, the point where the circle touches will either be along the bottom edge, when $\epsilon_1 = \lambda_{\textrm{min}}$, along the left-hand edge, when $\epsilon_0 = 1-\lambda_{\mathrm{max}}$, or at the bottom left-hand vertex, when $(\epsilon_0,\epsilon_1) = (1-\lambda_{\mathrm{max}},\lambda_{\textrm{min}})$. 

Second, if the smallest circle touching one of these edges (and not the corner), the edge will necessarily be tangent to the circle. 

Let us therefore consider first the case where the ellipse touches along the bottom edge of the rectangle. In this case, the point at which the two touch is where the tangent to the circle is horizontal. This will be at the point $(\epsilon_0^c, \epsilon_1^c+ \tfrac{c}{\sqrt{2}}) = (\tfrac{c}{2},\frac{c}{2}+\frac{c}{\sqrt{2}})$. That is, these points lie on the line
\begin{equation}
	\epsilon_1 = (1+\sqrt{2})\epsilon_0.
\end{equation}
This line must intersect the bottom of the rectangle. This will only happen if the intersection point is before the corner $(1~-~\lambda_{\rm max}, \lambda_{\rm min})$. The intersection occurs at $\epsilon_1 = \lambda_{\rm min}$, and so we obtain the region
\begin{equation}
	(1+\sqrt{2})(1-\lambda_{\rm max}) < \lambda_{\mathrm{min}}.
\end{equation}
The case where the circle touches along the left-hand edge is similar. We find that points with vertical tangent lie on the line $\epsilon_0 = (1+\sqrt{2})\epsilon_1$, the intersection occurs at $\epsilon_0 = 1-\lambda_{\mathrm{max}}$, and this will be above the corner when
\begin{equation}
(1+\sqrt{2})\lambda_{\textrm{min}} < 1-\lambda_{\mathrm{max}}.
\end{equation}
Finally, when neither of these cases hold, we see that the intersection will occur at the corner of the rectangle. %All of this is summarised in Fig.~\ref{f:optim}, and the result, 
Put together, leads to Table \ref{tab:1}, namely

\begin{equation}\label{e:3 regions}
	(\epsilon_0^*,\epsilon_1^*) = \begin{cases} 
		\left(1-\lambda_{\rm max}, (\sqrt{2}-1)(1-\lambda_{\rm max})\right) &\text{if}\quad \lambda_{\rm min} < (\sqrt{2}-1)(1-\lambda_{\rm max}),\\
				\left(1-\lambda_{\rm max}, \lambda_{\rm min}\right) &\text{if}\quad (\sqrt{2}-1)(1-\lambda_{\rm max}) \leq \lambda_{\rm min}\leq (\sqrt{2}+1)(1-\lambda_{\rm max}), \\
		\left((\sqrt{2}-1)\lambda_{\rm min}, \lambda_{\rm min}\right) &\text{if}\quad (\sqrt{2}+1)(1-\lambda_{\rm max}) < \lambda_{\rm min}.
	\end{cases}
\end{equation}
%\begin{figure}[t!]
%	\includegraphics[width=0.8\columnwidth]{optimisation.pdf}
%	\caption{\label{f:optimi} }	
%\end{figure}
\section{Generalised classical cloning can always distinguish a basis of states}
In this Appendix we will show that it is always possible to use a generalised classical cloning protocol to distinguish between a basis of states. As shown in the example in the main body in section \ref{ss:asympt}, all we in fact need to be able to do is distinguish a single basis state $\ket{i}$ (which we will take to be $\ket{0}$ without loss of generality) from the remaining basis states. That is, if we have $P(a|0) \neq P(a|i)$ for $i \neq 0$, even if all other $P(a|i)$ are equal (and hence do not allow us to distinguish the states $\ket{i}$, for $i \neq 0$, then we can still distinguish the entire basis, by a generalised protocol. In particular, the fact that $P(a|0) \neq P(a|i)$ means that with error which drops exponentially in the asymptotic limit of $N$ uses of the measurement, we can decide whether the state was $\ket{0}$ or not. 

Now, just as in the example in the main text, we could now consider simple pairwise permutations of the basis states, i.e. permuting $\ket{0}$ and $\ket{1}$, before applying the cloning procedure, and this would allow us then to determine whether the state was $\ket{1}$ or not. Using this basic idea, we can consider carrying out $d-1$ hypothesis tests in parallel by dividing the $N$ particles ($N-1$ ancillary particles and the incoming system) into $(d-1)$ groups (we choose $N$ to be divisible by $d-1$), and apply the following unitary
\begin{align}
    &U\ket{0}\ket{0}^{\otimes (N-1)} = \ket{e_0}^{\otimes N/(d-1)}\ket{e_1}^{\otimes N/(d-1)} \cdots \ket{e_{d-2}}^{\otimes N/(d-1)},\nonumber \\
    &U\ket{1}\ket{0}^{\otimes (N-1)} = \ket{e_1}^{\otimes N/(d-1)}\ket{e_0}^{\otimes N/(d-1)} \ket{e_{1}}^{\otimes (d-3)N/(d-1)},\nonumber \\
    &\hspace{7em}\vdots  \\
    &U\ket{d-2}\ket{0}^{\otimes (N-1)} = \ket{e_{d-2}}^{\otimes (d-2)N/(d-1)}\ket{e_0}^{\otimes N/(d-1)},\nonumber \\
    &U\ket{d-1}\ket{0}^{\otimes (N-1)} = \ket{e_{d-1}}^{\otimes N}.\nonumber
\end{align}
Considering the first group of $N/(d-1)$ particles, this can be used to determine whether the state is $\ket{0}$ or not. Considering the second group of $N/(d-1)$ particles, this can be used to determine whether the state is $\ket{1}$ or not. This continues until the last group of $N(d-1)$ particles, which is used to determine if the state is $\ket{d-2}$. If in each case we determine the state not to be one of the $d-1$ states checked for, then we can conclude that the state must have been the remaining state $\ket{d-1}$. This procedure thus allows us to determine the state. 

Concerning the errors made, each hypothesis test performed now uses $N/(d-1)$ particles, rather than the $N$ particles considered altogether previously. However, since $d$ is a constant, the probability of making an error still drops off exponentially in $N$, and hence it follows that altogether the error drops off exponentially in the number of uses $N$. 

The only case in which the above fails to work is if it is impossible to distinguish a single state $\ket{e_0}$ from the rest of a basis of states $\ket{e_i}$. That is, it must be the case that for all bases, the probability distributions $P(a|i) = \bra{e_i}M_a\ket{e_i}$ are independent of $i$.\footnote{Note that this doesn't mean that they have to be the same for all choices of basis $\ket{e_i}$, but just that for all bases, the probabilities are independent of the basis state.}. To see what this implies about the POVM elements $M_a$, let us consider that we use the eigenbasis $\ket{v_i^a}$ of $M_a$ as the basis. It follows that in order to fail, we have $\bra{v_i^a}M_a\ket{v_i^a} = \lambda_i^a = \lambda^a$ is constant for all $i$. That is, $M_a = \lambda^a \mathbb{I}$, each POVM element must be proportional to the identity operator. These are precisely the trivial measurements, which we have ruled out from the start, since it is clear that such measurements provide no information about the quantum state being measured, and therefore cannot be used to simulate any measurement. 

Altogether, this shows that the above procedure will always work for a non-trivial measurement; it must be possible to find a basis of states where at least one state is distinguishable from the remainder, and once this can be accomplished, the above shows that we can then in fact distinguish the entire basis of states by carefully applying permutations. 

We note finally that the above procedure is by no means optimal in general. In the general case, we would expect multiple states to be distinguished, and in this case, we can adapt the generalised cloning scheme to better make use of the information available. Nevertheless, it is important to realise that even with the weakest possible requirement (of only one state distinguishable), we can prove the desired result. 
\end{document}